\documentclass[12pt,preprint]{aastex}

\newcommand\arcpt{${{\lower3pt\hbox{$^{\prime\prime}$}}\atop{\raise4pt\hbox{.}}}$}
\newcommand\msun{$M_\odot$}
\newcommand\rsun{$R_\odot$}
\newcommand\sdeg{$^\circ$}
\def\kms{\ifmmode{\rm km\thinspace s^{-1}}\else km\thinspace s$^{-1}$\fi}

\slugcomment{to appear in the {\it Astrophysical Journal}}

\shorttitle{$\sigma^2$~CrB visual orbit}
\shortauthors{Raghavan et al.}

\begin{document}

\title{The Visual Orbit of the 1.1-day Spectroscopic Binary $\sigma^2$
       Coronae Borealis from Interferometry at the CHARA Array}

\author{Deepak Raghavan\altaffilmark{1,2}, 
        Harold A. McAlister\altaffilmark{1}, 
        Guillermo Torres\altaffilmark{3}, 
        David W. Latham\altaffilmark{3}, 
        Brian D. Mason\altaffilmark{4}, 
        Tabetha S. Boyajian\altaffilmark{1}, 
        Ellyn K. Baines\altaffilmark{1}, 
        Stephen J. Williams\altaffilmark{1}, 
        Theo A. ten Brummelaar\altaffilmark{5}, 
        Chris D. Farrington\altaffilmark{5}, 
        Stephen T. Ridgway\altaffilmark{6}, 
        Laszlo Sturmann\altaffilmark{5}, 
        Judit Sturmann\altaffilmark{5}, 
        Nils H. Turner\altaffilmark{5}}

\altaffiltext{1}{Center for High Angular Resolution Astronomy, Georgia
State University, P.O. Box 3969, Atlanta, GA 30302-3969}

\altaffiltext{2}{raghavan@chara.gsu.edu}

\altaffiltext{3}{Harvard-Smithsonian Center for Astrophysics, 60 Garden
Street, Cambridge, MA 02138}

\altaffiltext{4}{US Naval Observatory, 3450 Massachusetts Avenue NW, 
Washington DC 20392-5420}

\altaffiltext{5}{The CHARA Array, Mount Wilson Observatory, Mount Wilson,
CA 91023}

\altaffiltext{6}{National Optical Astronomy Observatory, P.O. Box 26732,
Tucson, AZ 85726-6732}

%%%%%%%%%%%%%%%%%%%%%%%%%%%%%%%%%%%%%%%%%%%%%%%%%%%%%%%%%%%%%%%%%%%%%%%%%%%%%%
\begin{abstract}
%%%%%%%%%%%%%%%%%%%%%%%%%%%%%%%%%%%%%%%%%%%%%%%%%%%%%%%%%%%%%%%%%%%%%%%%%%%%%%

We present an updated spectroscopic orbit and a new visual orbit for
the double-lined spectroscopic binary $\sigma^2$ Coronae Borealis
based on radial velocity measurements at the Oak Ridge Observatory in
Harvard, Massachusetts and interferometric visibility measurements at
the CHARA Array on Mount Wilson.  $\sigma^2$~CrB is composed of two
Sun-like stars of roughly equal mass in a circularized orbit with a
period of 1.14 days.  The long baselines of the CHARA Array have
allowed us to resolve the visual orbit for this pair, the shortest
period binary yet resolved interferometrically, enabling us to
determine component masses of 1.137 $\pm$ 0.037 \msun~and 1.090 $\pm$
0.036 \msun.  We have also estimated absolute $V$-band magnitudes of
$M_{\rm V}{\rm (primary)} = 4.35 \pm 0.02$ and $M_{\rm V}{\rm
(secondary)} = 4.74 \pm 0.02$.  A comparison with stellar evolution
models indicates a relatively young age of 1--3 Gyr, consistent with
the high Li abundance measured previously.  This pair is the central
component of a quintuple system, along with another similar-mass star,
$\sigma^1$~CrB, in a $\sim$ 730-year visual orbit, and a distant
M-dwarf binary, $\sigma$~CrB~C, at a projected separation of $\sim$
10$\arcmin$.  We also present differential proper motion evidence to
show that components C~\&~D (ADS 9979C~\&~D) listed for this system in
the Washington Double Star Catalog are optical alignments that are not
gravitationally bound to the $\sigma$~CrB system.

\end{abstract}


%%%%%%%%%%%%%%%%%%%%%%%%%%%%%%%%%%%%%%%%%%%%%%%%%%%%%%%%%%%%%%%%%%%%%%%%%%%%%%
\section{Introduction}
\label{sec:introduction}
%%%%%%%%%%%%%%%%%%%%%%%%%%%%%%%%%%%%%%%%%%%%%%%%%%%%%%%%%%%%%%%%%%%%%%%%%%%%%%

$\sigma$~CrB is a hierarchical multiple system 22 pc away.  Its
primary components, $\sigma^1$~CrB (HR 6064; HD 146362) and
$\sigma^2$~CrB (HR 6063; HD 146361), are in a visual orbit with a
preliminary period of $\sim$ 900 years \citep{Sca1979}, of which the
latter is an RS CVn binary with a circularized and synchronized orbit
of 1.139-day period \citep[][ SR03 hereafter]{Str2003}.  In addition
to these three solar-type stars, the Washington Double Star
Catalog\footnote{$http://ad.usno.navy.mil/wds/$} (WDS) lists three
additional components for this system.  WDS components C and D were
resolved 18$\arcsec$ away at 103\sdeg~in 1984 \citep{Pop1986} and
88$\arcsec$ away at 82\sdeg~in 1996 \citep{Cou1996}, respectively.  We
will show in \S\,\ref{sec:Opt} that both these components are optical
alignments that are not gravitationally bound to the $\sigma$~CrB
system.  Finally, WDS component E ($\sigma$~CrB~C, HIP 79551) which
was resolved 635$\arcsec$ away at 241\sdeg~in 1991 by
\textit{Hipparcos} \citep{HIP1997}, was identified as a
photocentric-motion binary by \citet{Hei1990}.  The parallax and
proper motion listed for this star in \citet{van2007}, the improved
\textit{Hipparcos} results based on a new reduction of the raw data,
match the corresponding measures for $\sigma^2$~CrB within the errors,
confirming a physical association.

SR03 presented photometric evidence in support of a rotation period of
1.157 $\pm$ 0.002 days for both components of $\sigma^2$~CrB, the
central pair of this system.  They explained the 0.017-day difference
between the rotation and orbital periods as differential surface
rotation. \citet{Bak1984} estimated an orbital inclination of 28\sdeg,
assuming component masses of 1.2 \msun~based on spectral types.  SR03
subsequently adopted this inclination to obtain component masses of
1.108 $\pm$ 0.004 \msun~and 1.080 $\pm$ 0.004 \msun, but these masses
are based on circular reasoning, and the errors are underestimated as
they ignore the uncertainty in inclination.  Several spectroscopic
orbits have been published for this pair (\citealt{Har1925};
\citealt{Bak1984}; \citealt{Duq1991}; SR03), enabling the
spectroscopic orbital elements to be well-constrained.  We present an
updated spectroscopic solution based on these prior data and our own
radial velocity measurements (\S\,\ref{sec:historical},
\S\,\ref{sec:SpecOrb}).  Our visual orbit leverages these
spectroscopic solutions and derives all orbital elements for this
binary (\S\,\ref{sec:visualorbit}), leading to accurate component
masses (\S\,\ref{sec:massest}).

This work utilizes a very precise parallax measure for this
radio-emitting binary obtained by \citet{Les1999} using Very Long
Baseline Interferometry (VLBI).  Their parallax of 43.93 $\pm$ 0.10
mas is about 10 times more precise than the \textit{Hipparcos} catalog
value of 46.11 $\pm$ 0.98 mas and 12 times more precise than the
\citet{van2007} measure of 47.35 $\pm$ 1.20 mas.  The
\citeauthor{Les1999} value is 2.2-$\sigma$ and 2.9-$\sigma$ lower than
the \textit{Hipparcos} and \citeauthor{van2007} measures,
respectively.  To check for systematic offsets, we compared the
parallaxes for all overlapping stars in these three sources.  While
the difference in parallax is most significant for $\sigma^2$~CrB, we
found no systematic differences.  Moreover, \citeauthor{Les1999}
performed statistical checks to verify the accuracy of their measure,
so we adopt their parallax to derive the physical parameters of the
component stars (\S\,\ref{sec:PhyPar}).

The Center for High Angular Resolution Astronomy (CHARA) Array's
unique capabilities, facilitated by the world's longest optical
interferometric baselines, have enabled a variety of astrophysical
studies \citep[e.g.,][]{McA2005, Bai2007, Mon2007}.  This work
utilizes the Array's longest baselines to resolve the 1.14-day
spectroscopic binary, the shortest period system yet resolved.  While
this is the first visual orbit determined using interferometric
visibilities measured with the CHARA Array, the technique described
here has regularly been employed for longer-period binaries using
other long-baseline interferometers \citep[e.g.,][]{Hum1993, Bod1999}.
The $\sigma^2$~CrB binary has a projected angular separation of about
1.1 mas in the sky, making it easily resolvable for the CHARA Array,
which has angular resolution capabilities in the $K'$ band down to
about 0.4 mas for binaries.

%%%%%%%%%%%%%%%%%%%%%%%%%%%%%%%%%%%%%%%%%%%%%%%%%%%%%%%%%%%%%%%%%%%%%%%%%%%%%%
\section{Spectroscopic Measurements}
\label{sec:spectroscopy}
%%%%%%%%%%%%%%%%%%%%%%%%%%%%%%%%%%%%%%%%%%%%%%%%%%%%%%%%%%%%%%%%%%%%%%%%%%%%%%

Spectroscopic observations of $\sigma^2$~CrB were conducted at the
Harvard-Smithsonian Center for Astrophysics (CfA) with an echelle
spectrograph on the 1.5m Wyeth reflector at the Oak Ridge Observatory
in the town of Harvard, Massachusetts.  A total of 46 usable spectra
were gathered from 1992 May to 1999 July, each of which covers a
single echelle order (45~\AA) centered at 5188.5~\AA\ and was recorded
using an intensified photon-counting Reticon detector
\citep[see][]{Lat1992}. The strongest lines in this window are those
of the \ion{Mg}{1}~b triplet. The resolving power of these
observations is $\lambda/\Delta\lambda \approx 35,\!000$, and the
nominal signal-to-noise ratios range from 21 to 94 per resolution
element of 8.5~\kms.

Radial velocities were obtained using the two-dimensional
cross-correlation algorithm TODCOR \citep{Zuc1994}. Templates for the
cross correlations were selected from an extensive library of
calculated spectra based on model atmospheres by R.\ L.\
Kurucz\footnote{Available at {\tt http://cfaku5.cfa.harvard.edu}.}
\citep[see also][]{Nor1994, Lat2002}. These calculated spectra cover a
wide range of effective temperatures ($T_{\rm eff}$), rotational
velocities ($v \sin i$ when seen in projection), surface gravities
($\log g$), and metallicities.  Experience has shown that radial
velocities are largely insensitive to the surface gravity and
metallicity adopted for the templates. Consequently, the optimum
template for each star was determined from extensive grids of
cross-correlations varying the temperature and rotational velocity,
seeking to maximize the average correlation weighted by the strength
of each exposure.  The results we obtain, adopting $\log g = 4.5$ and
solar metallicity\footnote{SR03 have reported a metallicity for
$\sigma^2$~CrB of [Fe/H] $= -0.37$ with an uncertainty no smaller than
0.1 dex, and \citet{Nor2004} reported the value [Fe/H] $= -0.24$ based
on Str\"omgren photometry.  Metallicity determinations for
double-lined spectroscopic binaries are particularly difficult, and
both of these estimates are likely to be affected at some level by the
double-lined nature of the system. However, the visual companion
($\sigma^1$~CrB) is apparently a single star, and has an accurate
spectroscopic abundance determination by \citet{Val2005} giving [Fe/H]
$= -0.06 \pm 0.03$, and another by \citet{Fuh2004} giving [Fe/H] $=
-0.064 \pm 0.068$. The near-solar metallicity from these
determinations is considered here to be more
reliable.\label{foot:metal}} for both stars, are $T_{\rm eff} =
6050$~K and $v \sin i = 26$~\kms\ for the primary, and $T_{\rm eff} =
5870$~K and $v \sin i = 26$~\kms\ for the secondary.  Estimated
uncertainties are 150~K and 1~\kms\ for the temperatures and projected
rotational velocities, respectively. Template parameters near these
values were selected for deriving the radial velocities.  Typical
uncertainties for the velocities are 1~\kms\ for both stars.

The stability of the zero-point of our velocity system was monitored
by means of exposures of the dusk and dawn sky, and small run-to-run
corrections were applied in the manner described by \citet{Lat1992}.
Additional corrections for systematics were applied to the velocities
as described by \citet{Lat1996} and \citet{Tor1997} to account for
residual blending effects. These corrections are based on simulations
with artificial composite spectra processed with TODCOR in the same
way as the real spectra. The final heliocentric velocities and their
1-$\sigma$ errors are listed in Table~\ref{tab:cfaRVs}, along with the
corresponding epochs of observation, $O\!-\!C$ residuals, and orbital
phase.

The light ratio between the components was estimated directly from the
spectra following \citet{Zuc1994}. After corrections for systematics
analogous to those described above, we obtain $\ell_{\rm s}/\ell_{\rm
p} = 0.67 \pm 0.02$ at the mean wavelength of our observations
(5188.5~\AA). Given that the stars have slightly different
temperatures, a small correction to the visual band was determined
from synthetic spectra integrated over the $V$ passband and the
spectral window of our observations. The corrected value is
($\ell_{\rm s}/\ell_{\rm p})_V = 0.70 \pm 0.02$.

The visual companion $\sigma^1$~CrB was also observed
spectroscopically at the CfA with the same instrumental setup. We
obtained 18 observations between 1996 June and 2004 August. The
stellar parameters were determined with a procedure similar to that
used for $\sigma^2$~CrB, and yielded $T_{\rm eff} = 5950 \pm 100$~K
and $v \sin i = 3 \pm 2$~\kms, for an adopted $\log g = 4.5$ and solar
metallicity (see Footnote~\ref{foot:metal}). Radial velocities were
obtained with standard cross-correlation techniques using a template
selected according to the above parameters. These measurements give an
average velocity of $-14.70 \pm 0.11$~\kms, with no significant
variation within the observational errors.  We use this radial
velocity to unambiguously determine the longitude of the ascending
node for the wider $\sigma^1\!-\!\sigma^2$~CrB visual orbit
(\S\,\ref{sec:outerVB}).

%%%%%%%%%%%%%%%%%%%%%%%%%%%%%%%%%%%%%%%%%%%%%%%%%%%%%%%%%%%%%%%%%%%%%%%%%%%%%%
\subsection{Historical Data Sets}
\label{sec:historical}
%%%%%%%%%%%%%%%%%%%%%%%%%%%%%%%%%%%%%%%%%%%%%%%%%%%%%%%%%%%%%%%%%%%%%%%%%%%%%%

In addition to our own, four other radial-velocity data sets have been
published in the literature (\citealt{Har1925}; \citealt{Bak1984};
\citealt{Duq1991}; SR03). Except for the more recent one, the older
data are generally of lower quality and contribute little to the mass
determinations, but they do extend the time coverage considerably (to
nearly 86 years, or $27,\!500$ orbital cycles) and can be used to
improve the orbital period.  Because of our concerns over possible
systematic differences among different data sets, particularly in the
velocity semi-amplitudes but also in the velocity zero points, we did
not simply merge all these observations together indiscriminately, but
instead we proceeded as follows. We considered all observations
simultaneously in a single least-squares orbital fit, imposing a
common period and epoch of maximum primary velocity in a circular
orbit, but we allowed each data set to have its own velocity
semi-amplitudes ($K_{\rm p}$, $K_{\rm s}$) as well as its own
systematic velocity zero-point offset relative to the reference frame
defined by the CfA observations.  Additionally, we included one more
adjustable parameter per set to account for possible systematic
differences between the primary and secondary velocities in each
group. These were statistically significant only in the observations
by SR03.  Relative weights for each data set were determined by
iterations from the RMS residual of the fit, separately for the
primary and secondary velocities. The resulting orbital period is $P =
1.139791423 \pm 0.000000080$ days, and the time of maximum primary
velocity nearest to the average date of the CfA observations is $T = 
2,\!450,\!127.61845 \pm 0.00020$ (HJD). We adopt this
ephemeris for the remainder of the paper.

%%%%%%%%%%%%%%%%%%%%%%%%%%%%%%%%%%%%%%%%%%%%%%%%%%%%%%%%%%%%%%%%%%%%%%%%%%%%%%
\section {Interferometric Measurements}
\label{sec:Interferometry}
%%%%%%%%%%%%%%%%%%%%%%%%%%%%%%%%%%%%%%%%%%%%%%%%%%%%%%%%%%%%%%%%%%%%%%%%%%%%%% 

Interferometric visibilities for $\sigma^2$~CrB were measured during
2007 May$-$July at the CHARA Array's six-element long-baseline
interferometer located on Mount Wilson, California \citep{ten2005}.
The Array uses the visible wavelengths 480--800 nm for tracking and
tip/tilt corrections, and the near-infrared $K'$ (2.13 $\mu$m) and $H$
(1.67 $\mu$m) bands for fringe detection.  The 26 visibility
measurements used in the final orbit determination, listed in
Table~\ref{tab:Visib}, were obtained in the $K'$ band on the S1-E1 and
S1-E2 two-telescope baselines spanning projected baselines of 268--331
meters.  The interference fringes were obtained using the pupil-plane
``CHARA Classic'' beam combiner.  While some of the data were obtained
via on-site observing at Mount Wilson, the bulk of the data were
gathered at the Arrington Remote Operations Center
\citep[AROC,][]{Fal2003} located at the Georgia State University
campus in Atlanta, Georgia.  Following the standard practice of
time-bracketed observations, we interleaved each target visibility
measurement with those of a calibrator star (HD 152598) in order to
remove instrumental and atmospheric effects.  For further details on
the observing practice and the data reduction process, refer to
\citet{McA2005}.

We selected HR 6279 (HD 152598), an F0V star offset from
$\sigma^2$~CrB by 8\fdg3, as the calibrator based on its small
estimated angular diameter and its apparent lack of any close
companions.  We obtained photometric measurements for this star in the
Johnson $UBV$ bands from \citet{Gre1985} and \citet{HIP1997}, and
$JHK_S$ bands from the Two Micron All Sky Survey\footnote{\tt
http://www.ipac.caltech.edu/2mass} (2MASS) and transformed them to
calibrated flux measurements using the methods described in
\citet{Col1996} and \citet{Coh2003}.  We then fitted these fluxes to
spectral energy distribution models\footnote{The model fluxes were
interpolated from the grid of models from R. L. Kurucz, available at
{\tt http://cfaku5.cfa.harvard.edu}}, yielding an angular diameter of
0.467 $\pm$ 0.013 mas for HD 152598, corresponding to $T_{\rm eff} =
7150$ K and $\log g = 4.3$.  This diameter estimate results in a
predicted calibrator visibility of $V_{\rm cal} = 0.858 \pm 0.008$ at
our longest baseline of 330 m, contributing roughly 1\% error to the
calibrated visibilities.  This error is included in our roughly 10\%
total visibility errors listed in Table~\ref{tab:Visib}, along with
the epoch of observation (at mid-exposure), the target star's
calibrated visibility, the predicted visibility for the best-fit
orbit, the $O\!-\!C$ visibility residual, the baseline projections
along East-West ($u$) and North-South ($v$) directions, and the hour
angle of the target.

%%%%%%%%%%%%%%%%%%%%%%%%%%%%%%%%%%%%%%%%%%%%%%%%%%%%%%%%%%%%%%%%%%%%%%%%%%%%%%
\section {Determination of the Orbit}
%%%%%%%%%%%%%%%%%%%%%%%%%%%%%%%%%%%%%%%%%%%%%%%%%%%%%%%%%%%%%%%%%%%%%%%%%%%%%% 

Consistent with prior evidence of a synchronized orbit (SR03), we
adopt a circular orbit ($e$ $\equiv$ 0, $\omega$ $\equiv$ 0) with the
orbital period ($P$) and epoch of nodal passage ($T$) from
\S\,\ref{sec:historical} for the spectroscopic and visual orbit
solutions presented below.

%%%%%%%%%%%%%%%%%%%%%%%%%%%%%%%%%%%%%%%%%%%%%%%%%%%%%%%%%%%%%%%%%%%%%%%%%%%%%%
\subsection {Spectroscopic Orbital Solutions}
\label{sec:SpecOrb}
%%%%%%%%%%%%%%%%%%%%%%%%%%%%%%%%%%%%%%%%%%%%%%%%%%%%%%%%%%%%%%%%%%%%%%%%%%%%%% 

\begin{figure}
\epsscale{1.0}\plotone{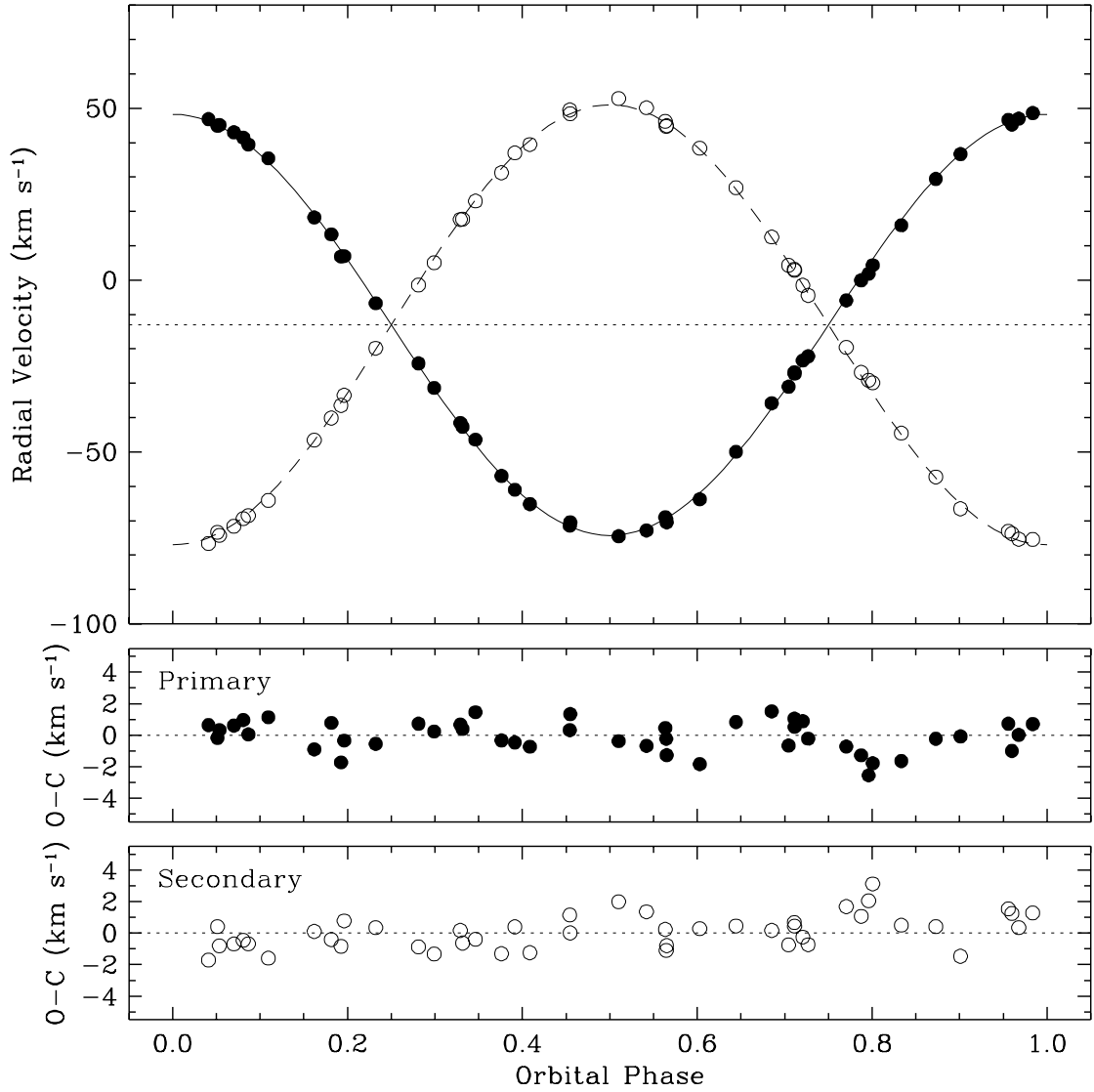}
\figcaption{Our radial velocities and the orbital fit for
$\sigma^2$~CrB (top panel) and the primary and secondary residuals
(bottom panels).  Filled circles represent the primary and open
circles represent the secondary component.  The corresponding orbital
elements are listed in Table~\ref{tab:SBele}.
\label{fig:cfa_orb}}
\end{figure}

\begin{figure}
\epsscale{1.0}\plotone{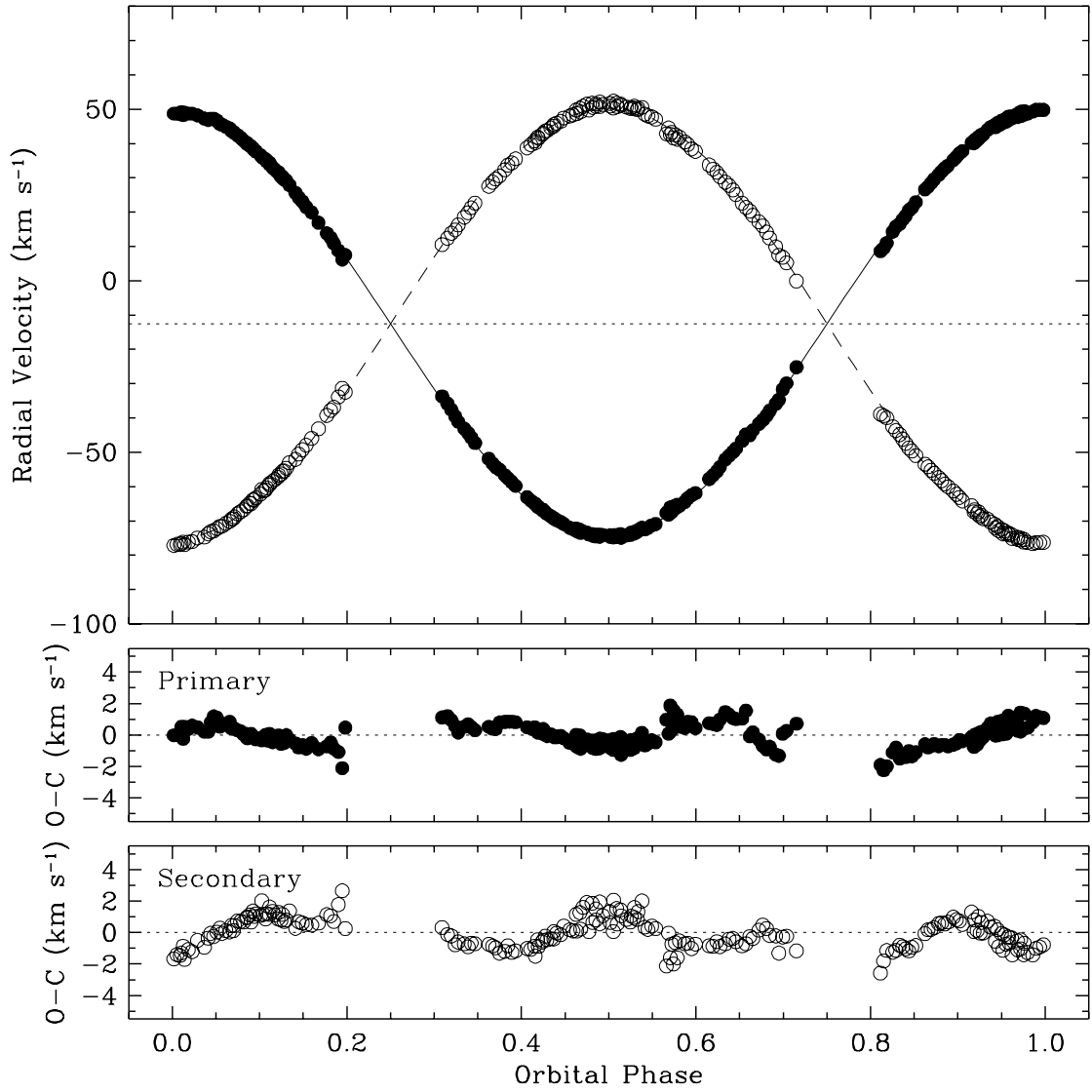}
\figcaption{Same as Figure \ref{fig:cfa_orb} but based on SR03 radial
velocities.
\label{fig:str_orb}}
\end{figure}

Our measured radial velocities enable us to derive the three remaining
spectroscopic orbital elements, namely, the center-of-mass velocity
($\gamma$) and the radial velocity semi-amplitudes of the primary and
secondary ($K_{\rm p}$ and $K_{\rm s}$, respectively).  To check for
consistency with prior efforts, we used the velocities published in
SR03 to derive a second orbital solution.  The calculated radial
velocities for the derived orbits are shown in
Figures~\ref{fig:cfa_orb} and~\ref{fig:str_orb} (solid and dashed
curves for the primary and secondary, respectively) along with the
measured radial velocities and residuals for the primary (filled
circles) and secondary (open circles).  The corresponding orbital
solutions are presented in Table~\ref{tab:SBele} along with the
related derived quantities.  For comparison purposes, we have also
included the values presented in SR03, which are consistent with our
orbit generated using their velocities.  However, the orbit obtained
using our velocities is statistically different from the one obtained
using SR03 velocities.  While the primary's velocity semi-amplitude
matches within the errors between these two solutions, the secondary's
differs by over 5-$\sigma$, resulting in a 4-$\sigma$ difference in
the mass ratios.

\begin{figure}
\epsscale{1.0}\plotone{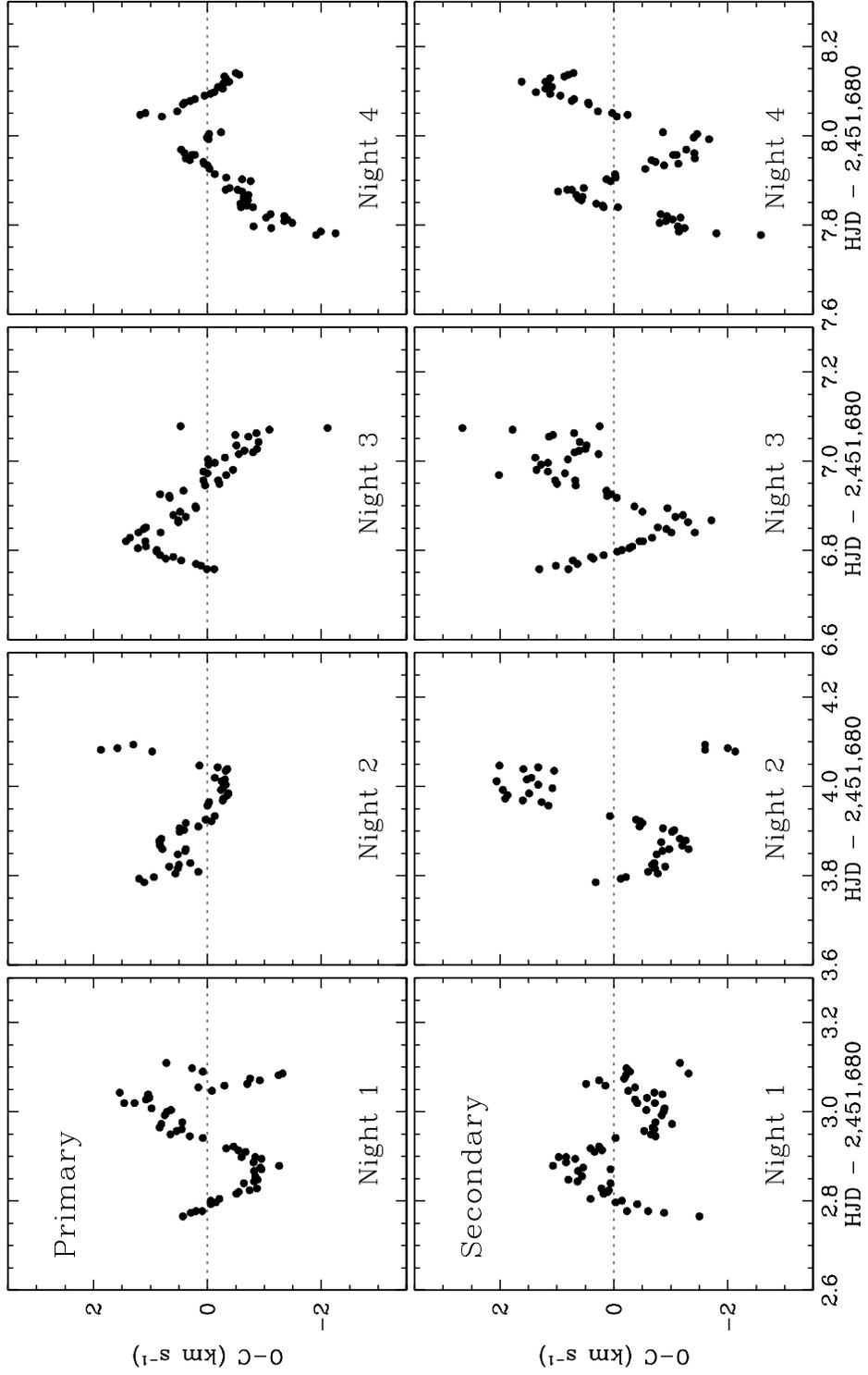}
\figcaption{Residuals for the individual nights' velocities from SR03.
\label{fig:str_resid}}
\end{figure}

One possible explanation of the difference in the orbital solutions
could be the velocity residuals for the orbit using SR03 data
(Figure~\ref{fig:str_orb}), which show an obvious pattern for both
components. Those observations were obtained on four nights over a
five-day period. To further examine these patterns, we display the
residuals for each of the four nights in Figure~\ref{fig:str_resid},
as a function of time. Clear trends are seen on each night, which are
different for the primary and secondary components and have
peak-to-peak excursions reaching 4~\kms\ in some cases, significantly
larger than the velocity errors of 0.1--1.2~\kms\ (SR03). On some but
not all nights, there appears to be a periodicity of roughly
0.20--0.25 days. The nature of these trends is unclear, particularly
because this periodicity is much shorter than either the orbital or the
rotational periods.  Instrumental effects seem unlikely, but an
explanation in terms of the considerable spottedness of both stars is
certainly a distinct possibility. The Doppler imaging maps produced by
SR03 show that both components display a very patchy distribution of
surface features covering the polar regions. Individual features
coming in and out of view as the stars rotate could easily be the
cause of the systematic effects observed in the radial velocities, and
the effects would not necessarily have to be the same on both stars,
just as observed.  Slight changes in the spots from one night to the
next could account for the different patterns seen in
Figure~\ref{fig:str_resid}. The relatively large amplitude of the
residual variations raises the concern that they may be affecting the
velocity semi-amplitudes of the orbit, depending on the phase at which
they occur.  We do not see such trends in the CfA data, perhaps
because our observations span a much longer time (more than 7 years,
and $\sim$2200 rotational cycles), allowing for spots to change and
average out these effects. We therefore proceed on the assumption that
possible systematic effects of this nature on $K_{\rm p}$ and $K_{\rm
s}$ are lessened in the CfA data.

%%%%%%%%%%%%%%%%%%%%%%%%%%%%%%%%%%%%%%%%%%%%%%%%%%%%%%%%%%%%%%%%%%%%%%%%%%%%%%
\section {The Visual Orbit Solution}
\label{sec:visualorbit}
%%%%%%%%%%%%%%%%%%%%%%%%%%%%%%%%%%%%%%%%%%%%%%%%%%%%%%%%%%%%%%%%%%%%%%%%%%%%%% 

The basic measured quantity from an interferometric observation is
\textit{visibility}, which evaluates the contrast in the fringe
pattern obtained by combining starlight wave fronts from multiple
apertures, filtered through a finite bandwidth.  For a single star of
angular diameter $\theta$, the interferometric visibility $V$ for a
uniform disk model is given by,
\begin{equation}
V = {2 J_1 (\pi B \theta / \lambda) \over \pi B \theta / \lambda},
\label{eq:Vsgl}
\end{equation}
where $J_1$ is the first-order Bessel function, $B$ is the projected
baseline length as seen by the star, and $\lambda$ is the observed
bandpass central wavelength.  The interferometric visibility for a
binary, where the individual stars have visibilities $V_{\rm p}$
(primary) and $V_{\rm s}$ (secondary) per equation (\ref{eq:Vsgl})
above, is given by
\begin{equation}
V = {\sqrt{(\beta^2 V_p^2 + V_s^2 + 2\beta V_pV_s \cos((2\pi /
\lambda)\bf{B\cdot s}) )} \over 1+\beta},
\label{eq:Vbin}
\end{equation}
where $\beta$ is the primary to secondary flux ratio, \textbf{B} is
the projected baseline vector as seen by the binary, and \textbf{s} is
the binary's angular separation vector in the plane of the sky.

Using our measured interferometric visibilities and the above
equations, we are able to augment the spectroscopic orbital solutions
to derive a visual orbit for $\sigma^2$~CrB.  Adopting the period and
epoch of nodal passage from \S\,\ref{sec:historical}, we now derive
the parameters that can only be determined astrometrically: angular
semimajor axis ($\alpha$); inclination ($i$); and, longitude of the
ascending node ($\Omega$).  We also treat the $K'$-band magnitude
difference as a free parameter in order to test evolutionary models.

For a circular orbit, the epoch of periastron passage ($T_{\rm 0}$) is
replaced by the epoch of ascending nodal passage ($T_{\rm node}$),
defined as the epoch of fastest secondary recession, in the visual
orbit equations \citep{Hei1978}.  Accordingly, we translate the $T$
value listed in \S\,\ref{sec:historical} by one-half of the orbital
period to determine the epoch of the ascending nodal passage as $T_{\rm
node} = 2,\!450,\!127.04855 \pm 0.00020$ (HJD) for use in our visual
orbit solution.  The 1-$\sigma$ errors of this and other adopted
parameters listed in Table~\ref{tab:VBele} have been propagated to our
error estimates for the derived parameters.

The angular diameters of the components are too small to be resolved
by our $K'$-band observations.  We therefore estimate these based on
the components' absolute magnitudes and temperatures as described
below. We first estimate the Johnson $V$-band magnitude of
$\sigma^2$~CrB using its Tycho-2 magnitudes of $B_{\rm T} = 6.262 \pm
0.014$ and $V_{\rm T} = 5.620 \pm 0.009$ and the relation $V_{\rm J} =
V_{\rm T} - 0.090(B_{\rm T} - V_{\rm T})$ from the Guide to the
Tycho-2 Catalog.  Then, using the $V$-band flux ratio from
\S\,\ref{sec:spectroscopy} and the \citet{Les1999} parallax, we obtain
absolute magnitudes of $M_{\rm V} = 4.35 \pm 0.02$ for the primary and
$M_{\rm V} = 4.74 \pm 0.02$ for the secondary.  These magnitudes lead
to linear radius estimates of 1.2\rsun~for the primary and 1.1\rsun~
for the secondary using the tabulation of stellar physical parameters
in \citet{Pop1980} and \citet{And1991}.  Finally, using the
\citet{Les1999} parallax, we adopt component angular diameters of
$\theta_{\rm p} = 0.50$ mas and $\theta_{\rm s} = 0.45$ mas,
propagating a 0.05 mas uncertainty in these values for deriving the
uncertainty of our orbital elements.  Diameter estimates using the
temperatures of the components from \S\,\ref{sec:spectroscopy} are
consistent with these values.

We conduct an exhaustive search of the parameter space for the unknown
parameters mentioned, namely, $\alpha$, $i$, $\Omega$, and $\Delta
K'$.  The orbital inclination is constrained by the $a \sin i$ from
spectroscopy, the free-parameter $\alpha$, and the \citet{Les1999}
parallax.  We impose this constraint during our exploration of the
parameter space along with its associated 1-$\sigma$ error.  We
explore the unknown parameters over many iterations, by randomly
selecting them between broad limits and using equation (\ref{eq:Vbin})
to evaluate the predicted binary visibility for the baseline and
binary positions at each observational epoch.  The orbital solution
presented here represents the parameter set with the minimum $\chi^2$
value when comparing the predicted and measured visibilities.

\begin{figure}
\epsscale{1.0}\plotone{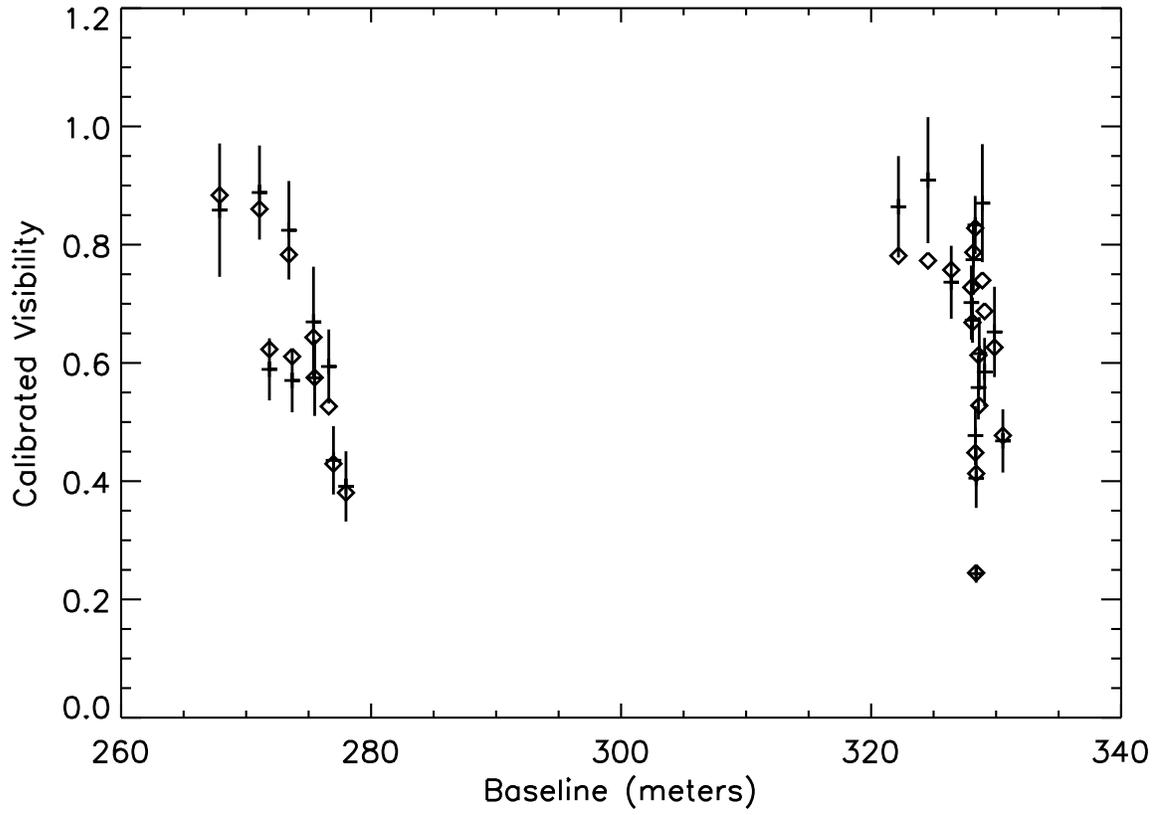}
\figcaption{Calibrated visibility measurements for $\sigma^2$~CrB
versus the projected baseline.  The plus signs are the calibrated
visibilities with vertical error bars, and the diamonds are the
calculated visibilities for the best-fit orbit.  Table~\ref{tab:Visib}
lists the numeric values corresponding to this plot.
\label{fig:Vfit}}
\end{figure}

\begin{figure}
\epsscale{1.0}\plotone{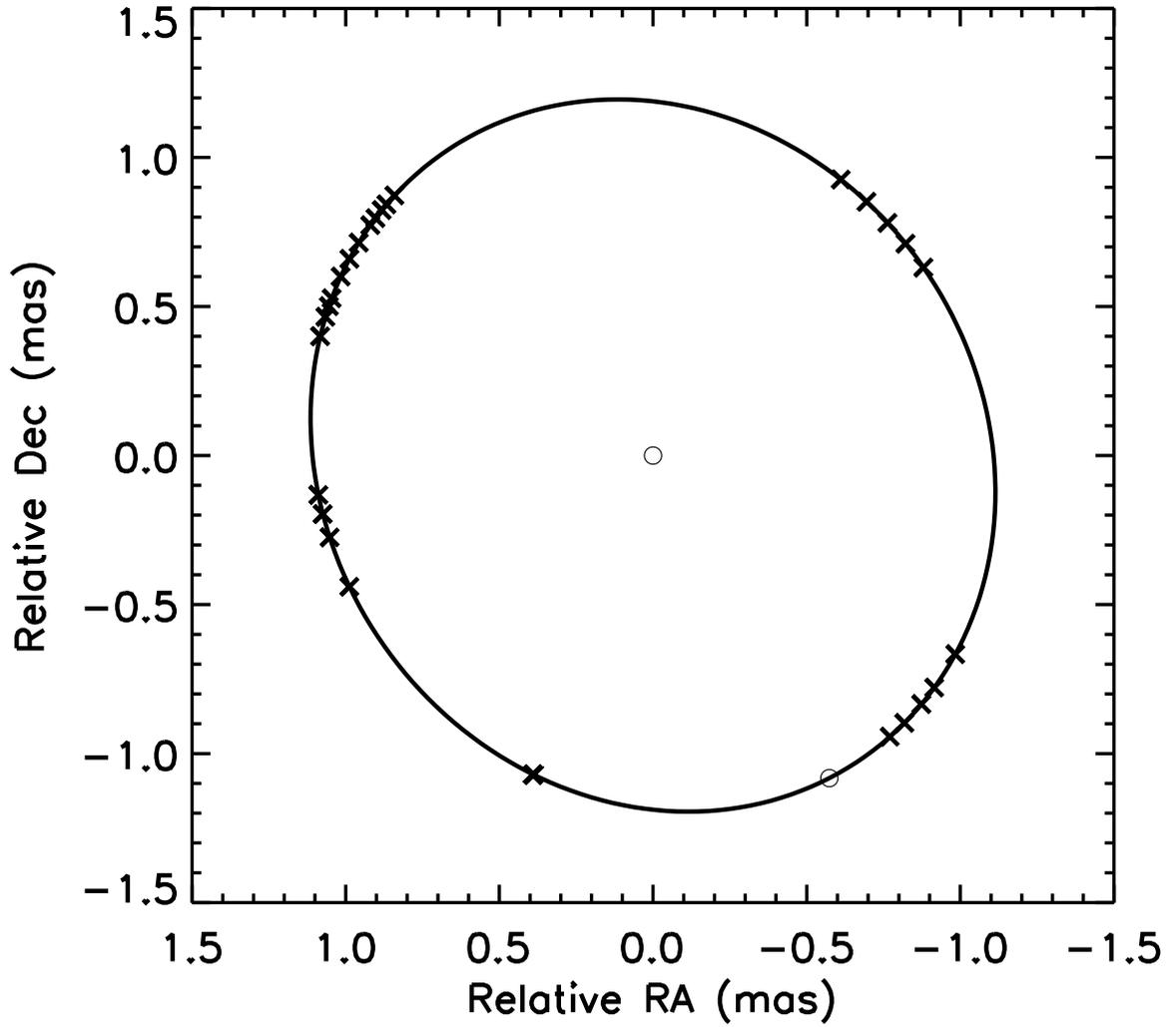}
\figcaption{The visual orbit of $\sigma^2$~CrB. Open circles mark the
positions of the two components at the epoch of ascending nodal
passage, and the X marks identify the secondary's calculated positions
at the epochs of visibility measurement.
\label{fig:OrbPlot}}
\end{figure}

Figure~\ref{fig:Vfit} shows the measured visibilities (plus signs)
with vertical error bars for each of the 26 observations, along with
the computed model visibilities (diamonds), and Table~\ref{tab:Visib}
lists the corresponding numerical values of the observed and model
visibilities along with the residuals of the fit.
Table~\ref{tab:VBele} summarizes the visual orbit parameters for
$\sigma^2$~CrB from our solution and Figure~\ref{fig:OrbPlot} plots
the visual orbit in the plane of the sky.  As seen in
Figure~\ref{fig:OrbPlot}, we have a reasonably good phase coverage
from our observations.

As mentioned in \S\,\ref{sec:SpecOrb}, star spots can create
systematic effects in the data obtained on this binary.  These effects
are especially significant for data obtained over a short time
baseline, as seen for the SR03 spectroscopic solution.  While our
interferometric data span 73 days, allowing for some averaging of
these effects, the bulk of the data used were obtained over 12 days,
justifying an exploration of this effect.  Specifically, the
separation between the stars derived from our visibility data would
represent the separation of the centers of light rather than that of
mass.  As discussed in \citet{Hum1994}, heavily-spotted stars will
incur a systematic shift in the center of light from rotational and
orbital motions, perhaps inducing an additional uncertainty in the
orbital elements derived. We assume a spot-induced change in the
angular semimajor axis of 2\% of the primary's diameter, or 0.01
mas. This is less that the uncertainty of our derived semimajor axis,
and at our baselines of 270--330 meters, translates to a 0.005--0.011
change in the visibility. While the uncertainties of our measured
visibilities are an order of magnitude larger than this, we ran a test
orbital fit by adding a 0.010 uncertainty to the visibility errors as
a root-sum-squared.  While, as expected, the $\chi^2$ of the fit
improved, the values and uncertainties of the derived parameters
remained unchanged, leading us to conclude that this effect, while
real, is too small to affect our results.

\begin{figure}
\epsscale{1.0}\plotone{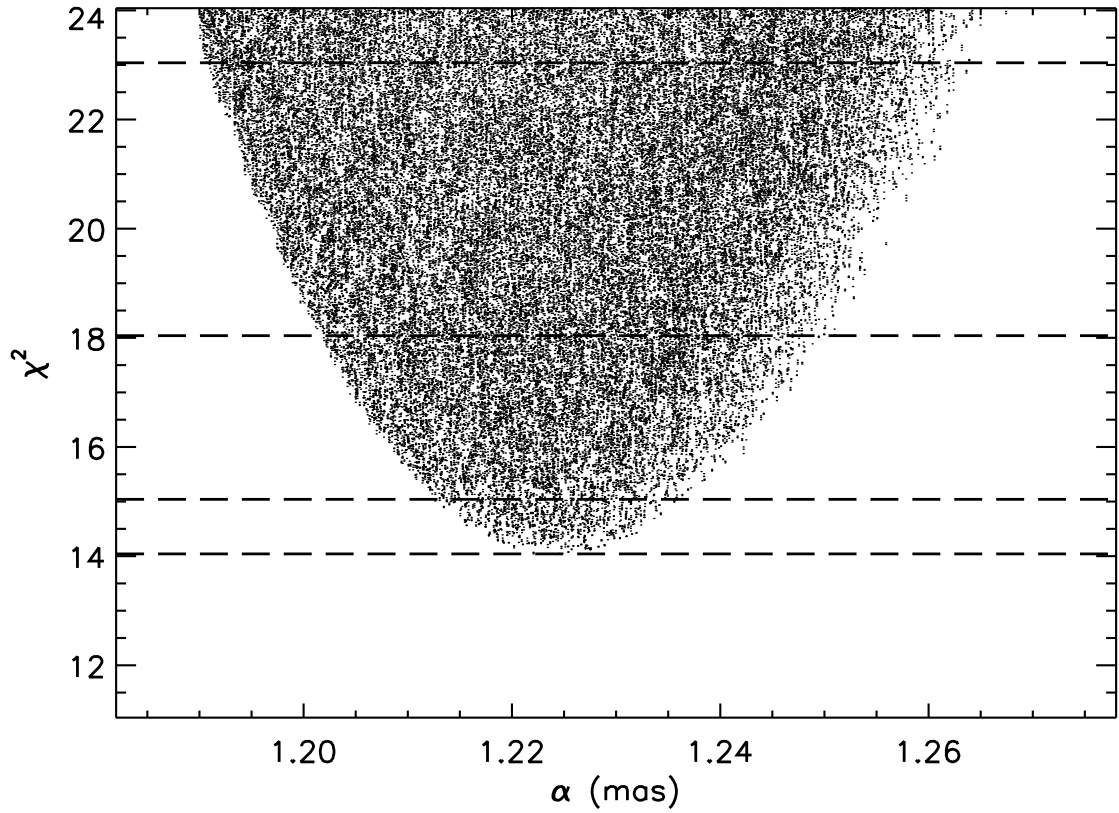}
\figcaption{$\chi^2$ distribution around the best-fit solution for the
angular semimajor axis ($\alpha$).  The bottom dashed line
corresponds to the minimum $\chi^2$ value, and the others mark a
deviation of 1, 4, and 9 units above the minimum, corresponding to 1,
2, and 3-$\sigma$ errors.
\label{fig:aerr}}
\end{figure}

\begin{figure}
\epsscale{1.0}\plotone{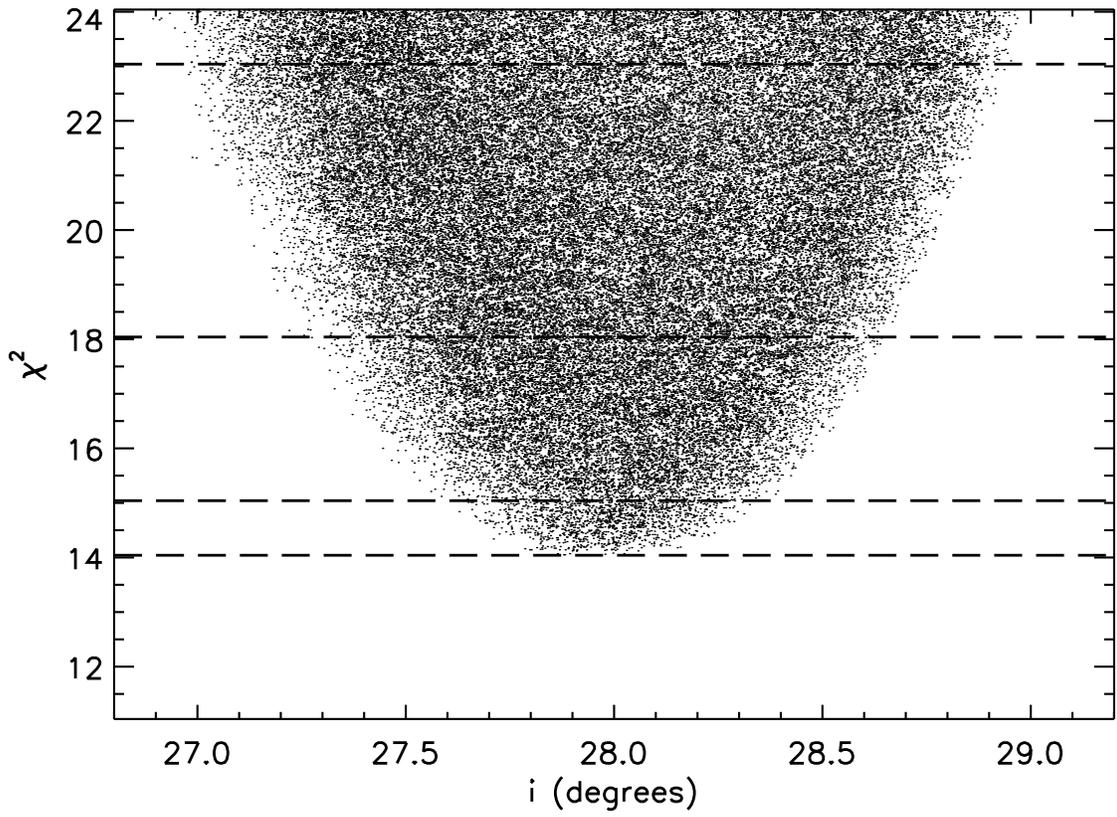}
\figcaption{Same as Figure~\ref{fig:aerr}, but for the orbital
inclination ($i$).
\label{fig:ierr}}
\end{figure}

\begin{figure}
\epsscale{1.0}\plotone{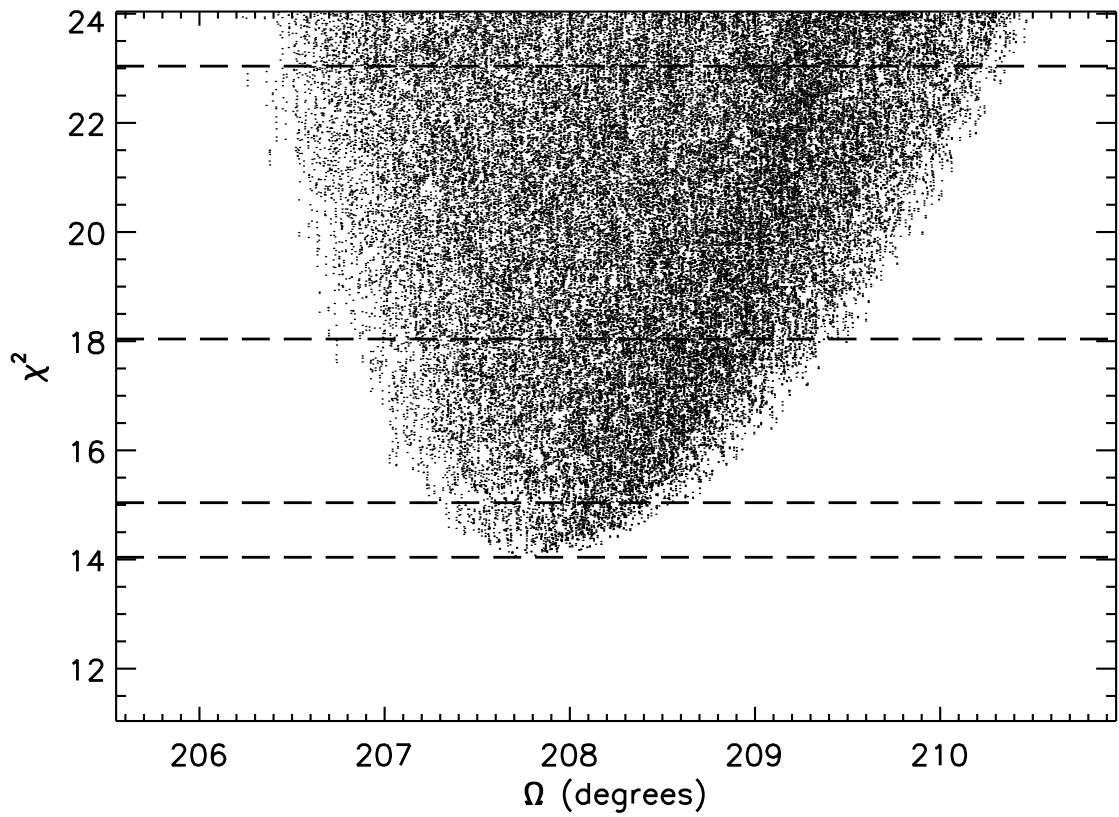}
\figcaption{Same as Figure~\ref{fig:aerr}, but for the longitude of
the ascending node ($\Omega$).
\label{fig:oerr}}
\end{figure}

\begin{figure}
\epsscale{1.0}\plotone{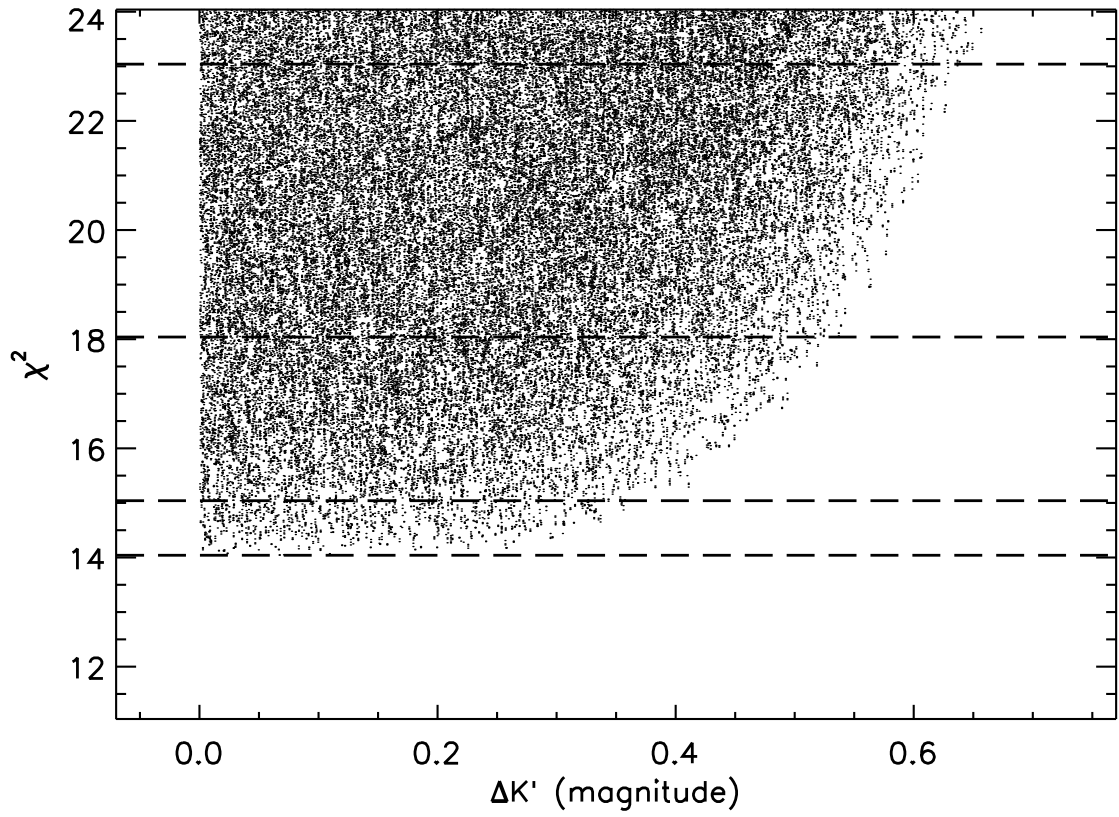}
\figcaption{Same as Figure~\ref{fig:aerr}, but for the $K'$-band
magnitude difference ($\Delta K'$).
\label{fig:dkerr}}
\end{figure}

We determine the 1, 2, and 3-$\sigma$ uncertainties of each visual
orbit parameter using a Monte Carlo simulation approach.  We compute
the orbital fit for $100,\!000$ iterations, where for each iteration, we
randomly select the adopted parameters within their respective
1-$\sigma$ intervals and the model parameters around their
corresponding best-fit solution, generating a multi-dimensional
$\chi^2$ ``surface''.  Then, we project this surface along each
parameter axis, resulting in the plots shown in Figures~\ref{fig:aerr}
to~\ref{fig:dkerr}.  The figures show the $\chi^2$ distribution around
the best-fit orbit and enable estimation of 1, 2, and 3-$\sigma$
errors for each parameter based on a $\chi^2$ deviation of 1, 4, and 9
units, respectively, from its minimum value.  The horizontal dashed
lines in the figures from bottom to top mark the minimum $\chi^2$
value and those corresponding to 1, 2, and 3-$\sigma$ errors, and
Table~\ref{tab:VBele} lists the corresponding numerical 1-$\sigma$
errors of the model parameters.

%%%%%%%%%%%%%%%%%%%%%%%%%%%%%%%%%%%%%%%%%%%%%%%%%%%%%%%%%%%%%%%%%%%%%%%%%%%%%%
\section {Physical Parameters}
\label{sec:PhyPar}
%%%%%%%%%%%%%%%%%%%%%%%%%%%%%%%%%%%%%%%%%%%%%%%%%%%%%%%%%%%%%%%%%%%%%%%%%%%%%%

\subsection{Component Mass Estimates}
\label{sec:massest}

Our angular semimajor axis obtained from interferometry translates
to 0.0279 $\pm$ 0.0003 AU or 5.99 $\pm$ 0.07 \rsun~using the
\citet{Les1999} parallax.  Newton's generalization of Kepler's Third
Law then yields a mass-sum of 2.227 $\pm$ 0.073 \msun~for the pair,
and using the mass ratio from our spectroscopic solution of 0.9586
$\pm$ 0.0047, we get individual component masses of 1.137 $\pm$ 0.037
\msun~and 1.090 $\pm$ 0.036 \msun~for the primary and secondary,
respectively.  As noted in \S\,\ref{sec:SpecOrb}, the SR03 velocities
yield a significantly different mass ratio of 0.9746 $\pm$ 0.0016, but
this 4-$\sigma$ difference is not enough to influence the mass
estimates significantly.  The uncertainty in our masses is dominated
by the cubed semimajor axis factor in estimating the mass sum,
resulting in about a 3\% uncertainty in mass-sum corresponding to a
1\% uncertainty in the semimajor axis.  The high precision of the mass
ratio from the spectroscopic solution results in final masses of 3\%
uncertainty as well. Component mass estimates using the SR03
velocities are 1.128 $\pm$ 0.037 and 1.099 $\pm$ 0.036, in excellent
agreement with the masses using our velocities.  These masses along
with other physical parameters derived are listed in
Table~\ref{tab:PhyPar}.

\subsection{Radii of the Components}

Assuming synchronous and co-aligned rotation of spherical components,
reasonable given the short orbital period and evidence from SR03 of
unevolved stars contained within their Roche limits, we can estimate
the component radii from the measured spectroscopic $v \sin i$.  As
mentioned in \S\,\ref{sec:spectroscopy}, our spectra yield $v \sin i =
26 \pm 1$ \kms\ for both the primary and secondary.  These values and
uncertainties are identical to those in SR03.  Using the inclination
from our visual orbit, and adopting the orbital period from
spectroscopy as the rotational period, we get identical component
radii of 1.244 $\pm$ 0.050 \rsun~for the primary and secondary.  This
translates to an angular diameter of 0.509 $\pm$ 0.020 mas for each
component using the \citet{Les1999} parallax, in excellent agreement
with our adopted diameter for the primary and a 1-$\sigma$ variance
for the secondary, given our associated 0.05-mas errors for these
values.  These radii estimates, along with the effective temperatures
from \S\,\ref{sec:spectroscopy} and the relation $L~\propto\ R^2\
T_{\rm eff}^4$, lead to a luminosity ratio of 0.89 $\pm$ 0.16.
Alternatively, using bolometric corrections from \citet{Flo1996} of
$BC_{\rm p} = -0.038 \pm 0.017$ and $BC_{\rm s} = -0.064 \pm 0.020$
corresponding to the components' effective temperatures, the $V$-band
flux ratio of 0.70 $\pm$ 0.02 from spectroscopy translates to a total
luminosity ratio of 0.68 $\pm$ 0.20, a 1-$\sigma$ variance from the
estimate above.  Conversely, our estimates of effective temperature
and luminosity ratio require a radius ratio of 0.88 $\pm$ 0.14, again
at a 1-$\sigma$ variance from the 1.00 $\pm$ 0.06 estimate from the
identical $v \sin i$ values of the components.

\subsection{Absolute Magnitudes and Ages}

We allowed the $K'$-band magnitude difference to be a free parameter
for our visual orbit fit, obtaining $\Delta K' = 0.19 \pm 0.19$,
consistent with the 0.18 estimate from the mass-luminosity relations
of \citet{Hen1993}\footnote{The relations from \citeauthor{Hen1993}
are for 0.5 \msun~$\le$ Mass $\le$ 1.0 ~\msun.  We consider it safe to
extrapolate out to our estimated masses of slightly larger than 1.0
\msun.}.  The uncertainty in $\Delta K'$ is large because visibility
measurements of nearly equal mass, and hence nearly equal brightness,
pairs are relatively insensitive to the magnitude difference of the
components \citep{Hum1998, Bod1999}.  Using equation (\ref{eq:Vbin}),
we have verified that a 10\% change in $\Delta K'$ for $\sigma^2$~CrB
results in only 0.1\% change in visibility.  This, along with the poor
quality $K$ magnitude listed in 2MASS (for $\sigma^2$~CrB, $K = 4.052
\pm 0.036$, but flagged as a very poor fit), thwart any attempts to
use these magnitudes for checking stellar evolution models.  However,
we can revert to $V$-band photometry to explore this topic.
 
\begin{figure}
\epsscale{1.0}\plotone{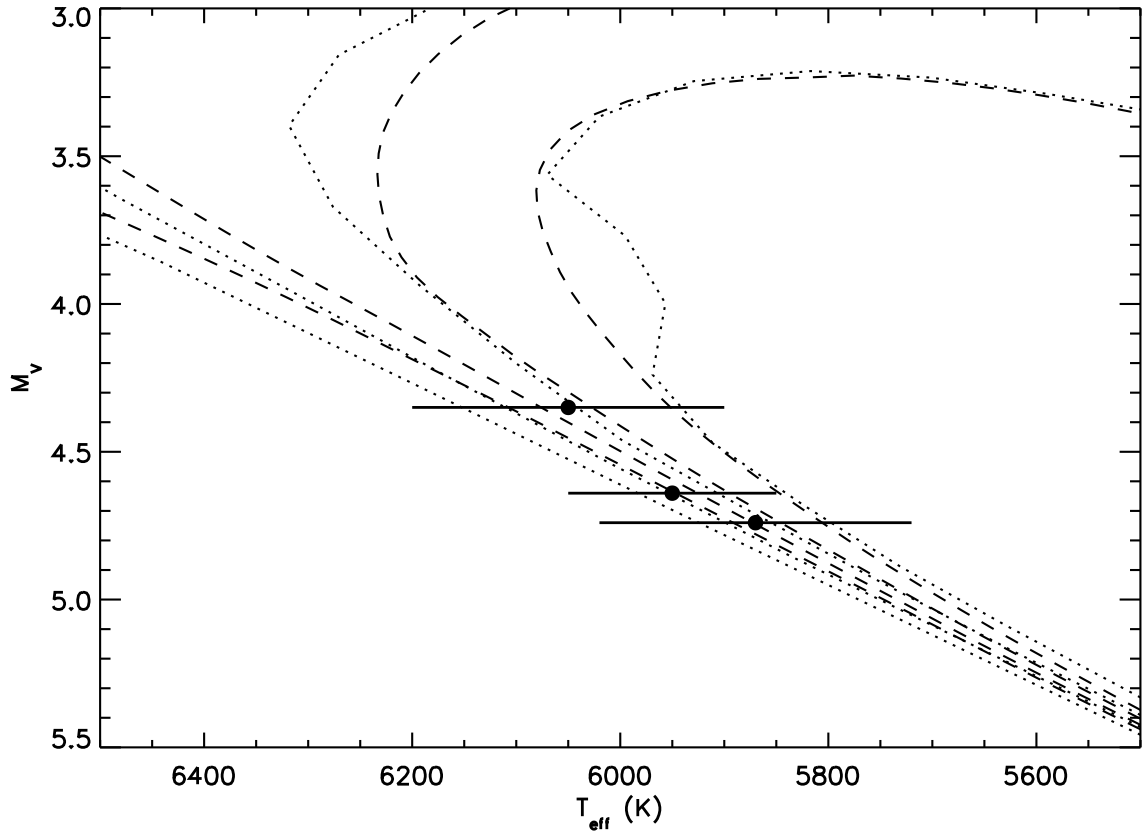} 
\figcaption{The position of the Sun-like components of $\sigma$~CrB on
the H-R diagram.  The points from top to bottom are $\sigma^2$~CrB
primary, $\sigma^1$~CrB, and $\sigma^2$~CrB secondary.  The isochrones
are from the Yonsei-Yale (dotted) and Victoria-Regina (dashed models)
for 0.5, 1.5, 3.0, and 5.0 Gyr ages (left to right) for solar
metallicity stars.
\label{fig:HRplot}}
\end{figure}

In \S\,\ref{sec:visualorbit}, we derived the absolute $V$-band
magnitudes of the components of $\sigma^2$~CrB as $M_{\rm V} = 4.35
\pm 0.02$ for the primary and $M_{\rm V} = 4.74 \pm 0.02$ for the
secondary.  For $\sigma^1$~CrB, we similarly use the Tycho-2
magnitudes and the \citet{Les1999} parallax to obtain $M_{\rm V} =
4.64 \pm 0.01$.  SR03 had a smaller magnitude difference for the
components of $\sigma^2$~CrB, and the corresponding results using
their spectroscopy are also included in Table~\ref{tab:PhyPar} along
with the values from their paper.  Figure~\ref{fig:HRplot} plots these
three stars on an H-R diagram using our magnitude and temperature
estimates, along with isochrones for 0.5, 1.5, 3.0, and 5.0 Gyr ages
(left to right) from the Yonsei-Yale isochrones
\citep[dotted,][]{Yi2001} and the Victoria-Regina stellar evolution
models \citep[dashed,][]{Van2006} for solar metallicity (see
Footnote~\ref{foot:metal}).

\citet{Wri2004} estimate an age of 1.8 Gyr for $\sigma^1$~CrB based on
chromospheric activity, and \citet{Val2005} estimate an age of 5.0 Gyr
from spectroscopy with limits of 2.9--7.8 Gyr based on 1-$\sigma$
changes to log$L$.  SR03 identify a much lower age, of a few times
10$^{\rm 7}$ years, by matching pre-main-sequence evolutionary tracks
and point to their higher Li abundance as supporting evidence.  While
abundance determinations in double-lined spectroscopic binaries are
particularly difficult and more prone to errors, the high Li abundance
of 2.60 $\pm$ 0.03 (SR03) for the slow-rotating single-lined companion
$\sigma^1$~CrB does argue for a young system.  Each point along the
isochrones plotted in Figure~\ref{fig:HRplot} corresponds to a
particular mass, allowing us to use our mass estimates for the
components of $\sigma^2$~CrB to further constrain the system's age.
Our mass, luminosity, and temperature estimates indicate an age for
this system of 0.5--1.5 Gyr, with a range of 0.1--3 Gyr permissible
within 1-$\sigma$ errors.

%%%%%%%%%%%%%%%%%%%%%%%%%%%%%%%%%%%%%%%%%%%%%%%%%%%%%%%%%%%%%%%%%%%%%%%%%%%%%%
\subsection {Mass Estimate of $\sigma^1$~CrB}
\label{sec:outerVB}
%%%%%%%%%%%%%%%%%%%%%%%%%%%%%%%%%%%%%%%%%%%%%%%%%%%%%%%%%%%%%%%%%%%%%%%%%%%%%% 

\begin{figure}
\epsscale{1.0}\plotone{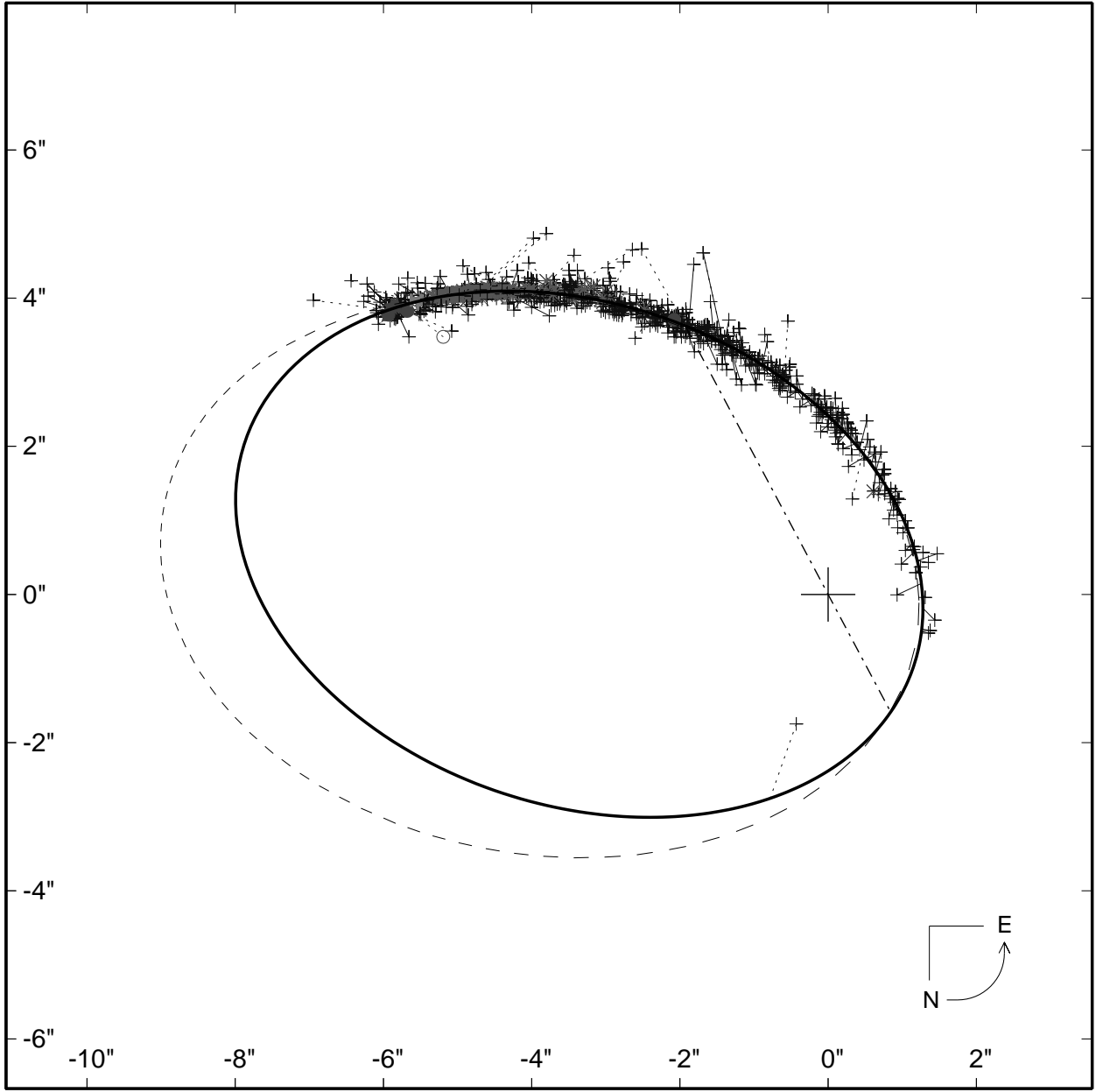}
\figcaption{The visual orbit of the wider
$\sigma^1\!-\!\sigma^2$~CrB~(AB) system based on all measures in
the WDS.  Plus signs indicate micrometric observations, asterisks
indicate photographic measures, open circles indicate eyepiece
interferometry, and filled circles represent speckle interferometry.
The solid curve is our orbit fit and the dashed curve is the
\citet{Sca1979} orbit.  $O\!-\!C$ lines connect each measure to its
predicted position along the orbit.  The big plus at the origin
indicates the position of the primary and the dot-dash line through it
is the line of nodes.  Scales are in arcseconds, and the curved arrow
at the lower right corner by the north and east direction indicators
shows the direction of orbital motion.
\label{fig:VBOab}}
\end{figure}

Our mass estimates for the components of $\sigma^2$~CrB allow us to
constrain the mass of the wider visual companion $\sigma^1$~CrB as
well.  \citet{Sca1979} presented an improved visual orbit for the AB
pair based on 886 observations spanning almost 200 years of
observation, yielding $P = 889$ years, $a = 5\farcs9$, $i = 31\fdg8$,
$e =0.76$, and $\Omega = 16\fdg9$.  However, he did not publish
uncertainties for these parameters, and given the long period, his
less than one-third phase coverage leads to only a preliminary orbital
solution, albeit one that convincingly shows orbital motion of the
pair.  He further uses parallaxes available to him to derive a
mass-sum for the AB system of 3.2 \msun.  We used all current WDS
observations, adding almost 200 observations since \citet{Sca1979}, to
update this orbit and obtain uncertainties for the parameters.  Our
visual orbit is presented in Figure~\ref{fig:VBOab}, along with the
\citeauthor{Sca1979} orbit for comparison, and  Table~\ref{tab:WideVis}
lists the derived orbital elements.  Adopting the \citet{Les1999}
parallax of the A component, we estimate a mass-sum of 3.2 $\pm$ 0.9
\msun, resulting in a B-component mass estimate of 1.0 \msun,
consistent with its spectral type of G1 V \citep{Gra2003}.
\citet{Val2005} estimate a mass of 0.77 $\pm$ 0.21 \msun\ based on
high-resolution spectroscopy, but we believe that they systematically
underestimate their uncertainty by overlooking the $\log e$ factor in
converting from uncertainty in $\log L$ to uncertainty in $L$.  Using
the $\log e$ factor, we followed their methods to obtain a mass
estimate of 0.77 $\pm$ 0.44 \msun.  The mass-error is dominated by the
uncertainty of the \citet{Gli1991} parallax used by \citet{Val2005}.
Adopting the higher precision \citet{Les1999} parallax of the primary,
we follow their method, and using the $\log e$ factor, get a mass
estimate of 0.78 $\pm$ 0.11 \msun.  This mass is too low for the
spectral type (as well as our own estimate of the effective
temperature; see \S\,\ref{sec:spectroscopy}) and the expectation from
the visual orbit.  A possible contamination of the secondary's
spectral type from the 7$\arcsec$ distant primary is unlikely, as
determined by Richard Gray at our request from new spectroscopic
observations (R.\ Gray 2008, private communication).

The inclination and longitude of the ascending node for this visual
orbit are similar to those of the inner ($\sigma^2$~CrB) orbit,
suggesting coplanarity.  For the outer visual orbit, we can use our
radial velocity estimate for $\sigma^1$~CrB, our derived systemic
velocity for $\sigma^2$~CrB, and the speckle observations to
unambiguously determine the longitude of the ascending node as $\Omega
= 28\fdg0 \pm 0\fdg5$.  Using the equation for the relative
inclination of the two orbits ($\phi$) from \citet{Fek1981}, we get
$\phi = 4\fdg7$ or 60\fdg3, given the 180\sdeg~ambiguity in $\Omega$
for the inner orbit, confirming coplanarity as a possibility.

%%%%%%%%%%%%%%%%%%%%%%%%%%%%%%%%%%%%%%%%%%%%%%%%%%%%%%%%%%%%%%%%%%%%%%%%%%%%%%
\section {The Wide Components: Optical or Physical?}
\label{sec:Opt}
%%%%%%%%%%%%%%%%%%%%%%%%%%%%%%%%%%%%%%%%%%%%%%%%%%%%%%%%%%%%%%%%%%%%%%%%%%%%%%

\begin{figure}
\epsscale{1.0}\plotone{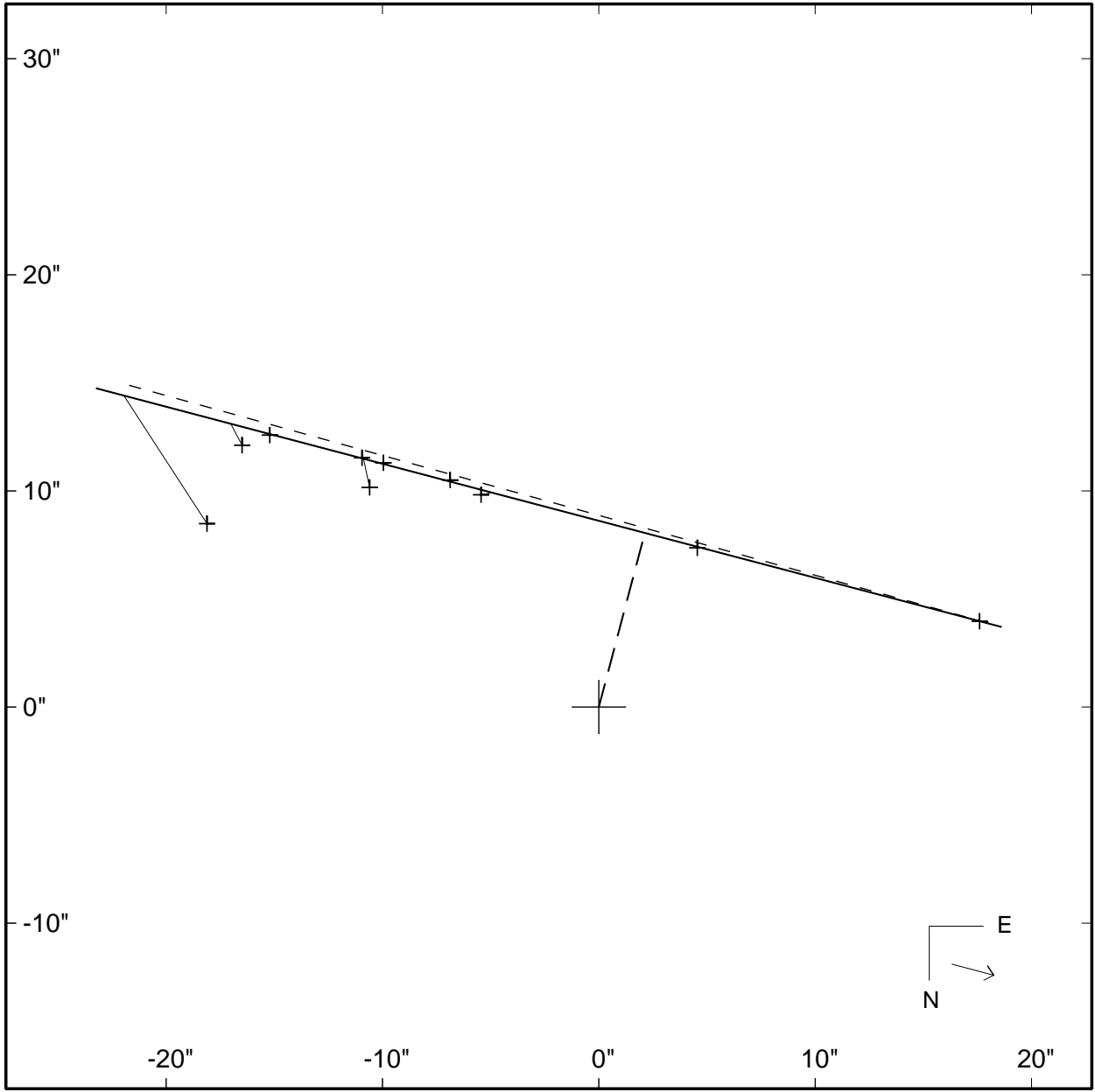}
\figcaption{Relative separation between $\sigma^2$~CrB and ADS 9979C
based on 10 resolutions of the pair from 1832 to 1984.  Plus signs
indicate micrometric observations.  $O\!-\!C$ lines connect each
measure to its predicted position along the linear fit (thick solid
line).  The thick dashed line is the predicted movement based on the
differential proper motions.  The long dashed line connected to the
origin indicates the predicted closest apparent position.  The scale
is in seconds of arc.  An arrow in the lower right corner by the north
and east direction indicators shows the direction of motion of the
star.
\label{fig:LinFitC}}
\end{figure}

\begin{figure}
\epsscale{1.0}\plotone{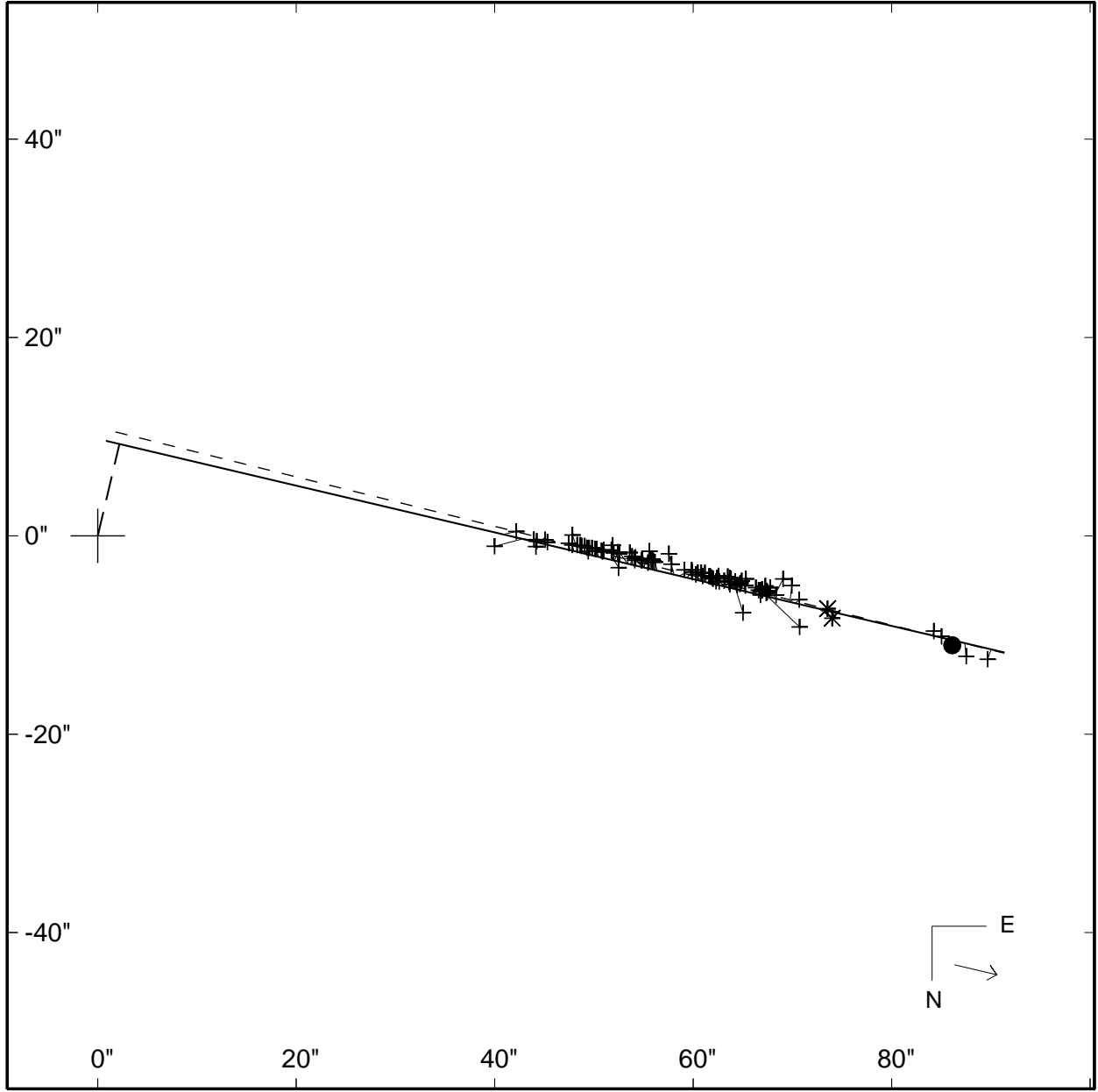}
\figcaption{Same as Figure~\ref{fig:LinFitC}, but for ADS 9979D based
on 106 resolutions of the pair from 1825 to 1996.  Asterisks indicate
photographic measures and filled circles represent Tycho measures.
\label{fig:LinFitD}}
\end{figure}

In addition to the three solar-type stars, the WDS lists three
additional components for $\sigma$~CrB.  We present evidence to show
that WDS components C and D are optical alignments, while component E,
itself a binary, is a physical association.  WDS component C (ADS
9979C), measured 18$\arcsec$ away at 103\sdeg~in 1984 \citep{Pop1986}
has a proper motion of $\mu_\alpha = -0\farcs016$ yr$^{-1}$ and
$\mu_\delta = -0\farcs015$ yr$^{-1}$ \citep{Jef1963}, significantly
different from that of $\sigma^2$~CrB of $\mu_\alpha = -0\farcs26364
\pm 0\farcs00091$ yr$^{-1}$ and $\mu_\delta = -0\farcs09259 \pm
0\farcs00129$ yr$^{-1}$ from \citet{van2007}.  Similarly, component D,
measured 88$\arcsec$ away at 82\sdeg~in 1996 \citep{Cou1996} and
clearly seen by us as a field star by blinking the multi-epoch STScI
Digitized Sky Survey\footnote{\tt
http://stdatu.stsci.edu/cgi-bin/dss\_$\!$ form} (DSS) images, has a
proper motion of $\mu_\alpha = +0\farcs004$ yr$^{-1}$ and $\mu_\delta
= -0\farcs017$ yr$^{-1}$ \citep{Jef1963}, again significantly
different from that of $\sigma^2$~CrB.  As a confirmation of the
optical alignment, we compare in Figures~\ref{fig:LinFitC}
and~\ref{fig:LinFitD} the observed separations of components C and D,
respectively, from the primary with the corresponding expected values
based on their proper motions.  The solid line is a linear fit to the
published measurements from the WDS and the dashed line is the
expected separation based on differential proper motion.  The
excellent agreement between the two lines for both components confirms
them as field stars.

WDS component E ($\sigma$~CrB~C, HIP 79551) is widely separated from
the primary at 635$\arcsec$, translating to a minimum physical
separation of over $14,\!000$ AU using the \citet{Les1999} parallax.
Despite its wide separation, this component appears to be physically
associated with $\sigma$~CrB based on its matching parallax of $\pi =
45.40 \pm 3.71$ mas and proper motion of $\mu_\alpha = -0\farcs26592
\pm 0\farcs00299$ yr$^{-1}$ and $\mu_\delta = -0\farcs08363 \pm
0\farcs00368$ yr$^{-1}$ \citep{van2007}.  While seemingly extreme for
gravitationally bound systems, physical association has been
demonstrated for pairs with separations out to $20,\!000$ AU
\citep[e.g.,][]{Lat1991, Pov1994}.  $\sigma$~CrB~C has a spectral
classification of M2.5V \citep{Rei1995}, apparent magnitude of $V =
12.24$ \citep{Bid1985}, and has itself been identified as a
photocentric-motion binary with an unseen companion of 0.1 \msun~in a
52-year orbit \citep{Hei1990}.  \citet{HIP1997} also identifies this
star as a binary of type `X' or stochastic solution, implying a
photocenter wobble for an unresolved star, but for which the
\textit{Hipparcos} data are not sufficient to derive an orbit.

%%%%%%%%%%%%%%%%%%%%%%%%%%%%%%%%%%%%%%%%%%%%%%%%%%%%%%%%%%%%%%%%%%%%%%%%%%%%%%
\section {Conclusion}
%%%%%%%%%%%%%%%%%%%%%%%%%%%%%%%%%%%%%%%%%%%%%%%%%%%%%%%%%%%%%%%%%%%%%%%%%%%%%%

\begin{figure}
\epsscale{1.0}\plotone{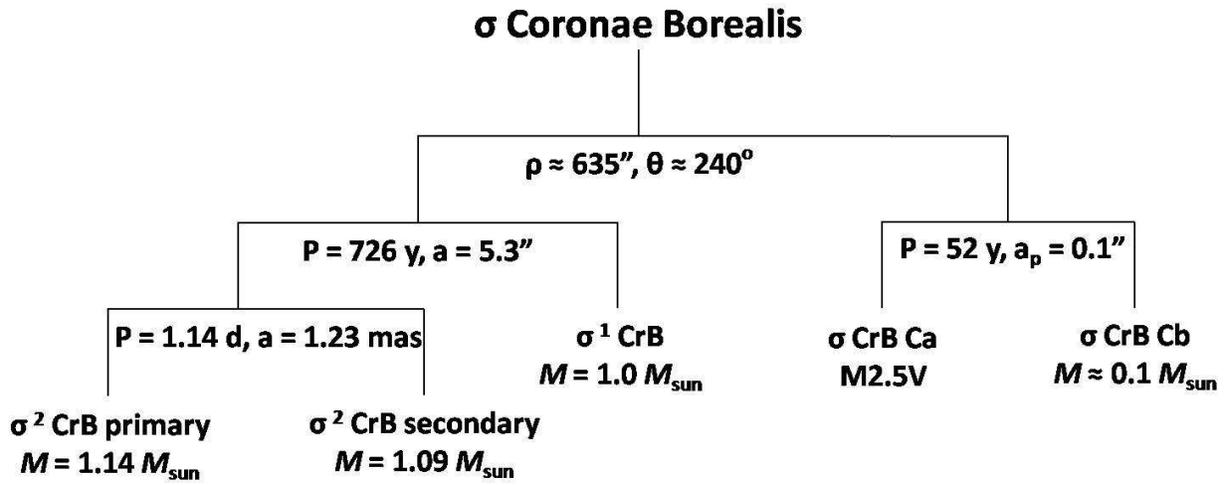}
\figcaption{Mobile diagram of $\sigma$~CrB and some of its properties.
The Ca-Cb pair is WDS component E, while WDS components C~\&~D are not
gravitationally bound to the $\sigma$~CrB system (see Figures
~\ref{fig:LinFitC} and \ref{fig:LinFitD}, and text in
\S\,\ref{sec:Opt}).  $a_p$ for the Ca-Cb pair is the photocentric
semimajor axis.
\label{fig:MobDia}}
\end{figure}

Augmenting our radial velocity measurements with published values, we
obtain a coverage of nearly 86 years or $27,\!500$ orbital cycles,
resulting in a very precise ephemeris of $P = 1.139791423 \pm
0.000000080$ days and $T = 2,\!450,\!127.61845 \pm 0.00020$ (HJD) and
a robust spectroscopic orbit for $\sigma^2$~CrB.  Using the CHARA
Array, we have resolved this 1.14-day spectroscopic binary, the
shortest period system yet resolved, and derived its visual orbit.
The resulting component masses are 1.137 $\pm$ 0.037 \msun~and 1.090
$\pm$ 0.036 \msun~for the primary and secondary, respectively.  Our
spectroscopy supports prior efforts in estimating the same $v \sin i$
values for both components, which assuming a synchronized, co-aligned
rotation, results in equal radii of 1.244 $\pm$ 0.050 \rsun~for both
components.  The corresponding radius ratio is consistent within
1-$\sigma$ with its estimate using the components' temperatures and
flux ratio from spectroscopy.  We have also shown that this binary
resides in a hierarchical quintuple system, composed of three close
Sun-like stars and a wide M-dwarf binary.  The wider visual orbit
companion, $\sigma^1$~CrB, is about 7$\arcsec$ away in a 726-year
visual orbit with $i = 32\fdg3$, which appears to be coplanar with the
inner orbit.  A comparison of the mass and absolute magnitude
estimates of $\sigma^1$~CrB and $\sigma^2$~CrB with current stellar
evolution models indicates a young age for the system of 1--3 Gyr,
consistent with the relatively high Li abundance previously measured.
Finally, the widest member of this system is an M-dwarf binary,
$\sigma$~CrB~C, at a minimum separation of $14,\!000$ AU.
Figure~\ref{fig:MobDia} depicts the system's hierarchy in a pictorial
form.

%%%%%%%%%%%%%%%%%%%%%%%%%%%%%%%%%%%%%%%%%%%%%%%%%%%%%%%%%%%%%%%%%%%%%%%%%%%%%%
\acknowledgments
%%%%%%%%%%%%%%%%%%%%%%%%%%%%%%%%%%%%%%%%%%%%%%%%%%%%%%%%%%%%%%%%%%%%%%%%%%%%%%

We thank Andy Boden and Doug Gies for their many useful suggestions
that improved the quality of this work, and Richard Gray for making
new observations at our request to confirm the spectral typing of the
components.  The CfA spectroscopic observations of $\sigma^1$~CrB and
$\sigma^2$~CrB used in this paper were obtained with the help of J.\
Caruso, R.\ P.\ Stefanik, and J.\ Zajac.  We also thank the CHARA
Array operator P. J. Goldfinger for obtaining some of the data used
here and for her able assistance of remote operations of the Array
from AROC.  Research at the CHARA Array is supported by the College of
Arts and Sciences at Georgia State University and by the National
Science Foundation through NSF grant AST-0606958.  GT acknowledges
partial support for this work from NSF grant AST-0708229 and NASA's
MASSIF SIM Key Project (BLF57-04).  This research has made use of the
SIMBAD literature database, operated at CDS, Strasbourg, France, and
of NASA's Astrophysics Data System.  This effort used multi-epoch
images from the Digitized Sky Survey, which was produced at the Space
Telescope Science Institute under U.S. Government grant NAG W-2166.
This publication also made use of data products from the Two Micron
All Sky Survey (2MASS), which is a joint project of the University of
Massachusetts and the Infrared Processing and Analysis
Center/California Institute of Technology, funded by the National
Aeronautics and Space Administration and the National Science
Foundation.

%%%%%%%%%%%%%%%%%%%%%%%%%%%%%%%%%%%% REFS %%%%%%%%%%%%%%%%%%%%%%%%%%%%%%%%%%%%

\clearpage

%%%%%%%%%%%%%%% TABLE: CfA Radial  Velocities%%%%%%%%%%%%%%%%%%%

%\voffset30pt{}
\begin{deluxetable}{cccccccc}
\tabletypesize{\footnotesize}
\tablecaption{Radial Velocities of $\sigma^2$~CrB
\label{tab:cfaRVs}}
\tablewidth{0pt}

\tablehead{%\vspace{-25pt} \\
           \colhead{HJD}&
           \colhead{$RV_{\rm p}$}&
           \colhead{$RV_{\rm s}$}&
           \colhead{$\sigma_{RV_{\rm p}}$}&
           \colhead{$\sigma_{RV_{\rm s}}$}&
           \colhead{$(O-C)_{\rm p}$}&
           \colhead{$(O-C)_{\rm s}$}&
           \colhead{Orbital} \\

           \colhead{($2,\!400,\!000$+)}&
           \colhead{(\kms)}&
           \colhead{(\kms)}&
           \colhead{(\kms)}&
           \colhead{(\kms)}&
           \colhead{(\kms)}&
           \colhead{(\kms)}&
           \colhead{Phase}}

\startdata

48764.6474 & \phs\phn6.88 &     $-$36.45 & 2.68 & 2.84 &  $-$1.72 &  $-$0.87 & 0.193 \\
48781.6495 &    \phs35.46 &     $-$64.08 & 2.99 & 3.16 & \phs1.15 &  $-$1.68 & 0.109 \\
48810.6618 &     $-$69.00 &    \phs46.22 & 1.16 & 1.23 & \phs0.47 & \phs0.37 & 0.564 \\
48813.6236 &    \phs18.25 &     $-$46.52 & 1.19 & 1.26 &  $-$0.89 & \phs0.06 & 0.162 \\
48820.6185 &     $-$31.35 & \phs\phn5.07 & 1.61 & 1.71 & \phs0.24 &  $-$1.27 & 0.299 \\
48822.6494 &    \phs41.46 &     $-$69.41 & 1.32 & 1.40 & \phs0.97 &  $-$0.55 & 0.081 \\
48826.5581 &     $-$74.53 &    \phs52.87 & 1.19 & 1.26 &  $-$0.38 & \phs2.13 & 0.510 \\
48828.6849 &     $-$56.96 &    \phs31.25 & 1.37 & 1.45 &  $-$0.33 &  $-$1.21 & 0.376 \\
48838.5942 &    \phs43.01 &     $-$71.62 & 1.15 & 1.22 & \phs0.62 &  $-$0.79 & 0.070 \\
50258.6759 &    \phs48.63 &     $-$75.42 & 1.43 & 1.51 & \phs0.73 & \phs1.17 & 0.984 \\
50260.6371 &     $-$31.00 & \phs\phn4.33 & 0.85 & 0.90 &  $-$0.66 &  $-$0.71 & 0.704 \\
50263.6316 &     $-$42.68 &    \phs17.76 & 0.83 & 0.88 & \phs0.40 &  $-$0.56 & 0.332 \\
50266.6225 &    \phs46.61 &     $-$73.03 & 0.99 & 1.04 & \phs0.74 & \phs1.43 & 0.956 \\
50269.7633 &     $-$27.25 & \phs\phn2.84 & 0.99 & 1.05 & \phs0.53 & \phs0.47 & 0.711 \\
50271.6269 &     $-$46.41 &    \phs23.01 & 0.95 & 1.01 & \phs1.46 &  $-$0.31 & 0.346 \\
50275.6464 &    \phs29.47 &     $-$57.26 & 0.97 & 1.03 &  $-$0.22 & \phs0.33 & 0.873 \\
50285.6440 &     $-$49.95 &    \phs26.91 & 0.90 & 0.95 & \phs0.84 & \phs0.54 & 0.644 \\
50287.6352 &     $-$60.98 &    \phs37.03 & 0.89 & 0.94 &  $-$0.45 & \phs0.51 & 0.391 \\
50292.5697 &     $-$23.39 &  \phn$-$1.49 & 1.02 & 1.08 & \phs0.90 &  $-$0.22 & 0.721 \\
50295.6335 &     $-$65.17 &    \phs39.49 & 0.79 & 0.83 &  $-$0.72 &  $-$1.13 & 0.409 \\
50298.5502 &    \phs46.99 &     $-$75.36 & 0.71 & 0.75 & \phs0.03 & \phs0.24 & 0.968 \\
50300.5553 &     $-$22.15 &  \phn$-$4.43 & 0.80 & 0.85 &  $-$0.21 &  $-$0.70 & 0.727 \\
50302.6499 &     $-$69.55 &    \phs44.72 & 0.84 & 0.89 &  $-$0.23 &  $-$0.98 & 0.565 \\
50346.5051 &    \phs46.86 &     $-$76.63 & 0.92 & 0.97 & \phs0.65 &  $-$1.81 & 0.041 \\
50348.5107 & \phs\phn4.35 &     $-$29.89 & 0.99 & 1.04 &  $-$1.77 & \phs3.10 & 0.801 \\
50350.5649 &     $-$63.76 &    \phs38.37 & 0.81 & 0.86 &  $-$1.83 & \phs0.39 & 0.603 \\
50352.4779 &     $-$24.23 & \phn $-$1.41 & 0.79 & 0.84 & \phs0.74 &  $-$0.84 & 0.281 \\
50356.4742 &  \phn$-$0.04 &     $-$26.85 & 0.79 & 0.84 &  $-$1.27 & \phs1.05 & 0.787 \\
50358.4740 &     $-$72.84 &    \phs50.15 & 0.77 & 0.81 &  $-$0.68 & \phs1.49 & 0.542 \\
50361.4826 &    \phs13.31 &     $-$40.12 & 0.80 & 0.85 & \phs0.79 &  $-$0.45 & 0.182 \\
50364.4624 & \phs\phn1.84 &     $-$29.15 & 0.86 & 0.91 &  $-$2.54 & \phs2.04 & 0.796 \\
50374.4574 &     $-$70.50 &    \phs44.94 & 0.85 & 0.90 &  $-$1.26 &  $-$0.67 & 0.565 \\
50379.4665 &    \phs45.29 &     $-$73.75 & 0.82 & 0.87 &  $-$0.99 & \phs1.14 & 0.960 \\
50383.4500 &     $-$70.47 &    \phs48.43 & 0.84 & 0.89 & \phs1.34 & \phs0.13 & 0.455 \\
50385.4760 &  \phn$-$6.74 &     $-$19.80 & 0.81 & 0.86 &  $-$0.54 & \phs0.35 & 0.232 \\
50388.4407 &    \phs15.96 &     $-$44.52 & 0.92 & 0.97 &  $-$1.63 & \phs0.45 & 0.833 \\
50391.4280 &     $-$71.44 &    \phs49.52 & 0.81 & 0.86 & \phs0.32 & \phs1.28 & 0.454 \\
50590.7488 &     $-$41.53 &    \phs17.65 & 0.98 & 1.04 & \phs0.68 & \phs0.24 & 0.329 \\
50619.6791 &     $-$26.78 & \phs\phn3.14 & 1.05 & 1.11 & \phs1.06 & \phs0.72 & 0.711 \\
50846.9255 &    \phs39.45 &     $-$68.48 & 0.90 & 0.95 & \phs0.06 &  $-$0.78 & 0.087 \\
51216.9001 &     $-$35.81 &    \phs12.55 & 1.98 & 2.09 & \phs1.52 & \phs0.23 & 0.685 \\
51246.7808 &    \phs36.69 &     $-$66.52 & 2.01 & 2.13 &  $-$0.06 &  $-$1.57 & 0.901 \\
51279.6859 &  \phn$-$5.90 &     $-$19.52 & 2.51 & 2.65 &  $-$0.71 & \phs1.68 & 0.770 \\
51341.7199 & \phs\phn6.97 &     $-$33.48 & 1.77 & 1.87 &  $-$0.33 & \phs0.75 & 0.196 \\
51374.6086 &    \phs44.93 &     $-$73.34 & 2.01 & 2.12 &  $-$0.16 & \phs0.31 & 0.051 \\
51374.6112 &    \phs45.14 &     $-$74.26 & 3.08 & 3.26 & \phs0.34 &  $-$0.91 & 0.054 \\

\enddata

\end{deluxetable}

\clearpage

%%%%%%%%%%%%%%%%% TABLE: Observed & Model Visib %%%%%%%%%%%%%%%%%%%%

%\voffset30pt{}
\begin{deluxetable}{cccccccc}
\tabletypesize{\footnotesize}
\tablecaption{Interferometric Visibilities for $\sigma^2$~CrB
\label{tab:Visib}}
\tablewidth{0pt}

\tablehead{%\vspace{-25pt} \\
           \colhead{HJD}&
           \colhead{}&
           \colhead{}&
           \colhead{}&
           \colhead{}&
           \colhead{$u$}&
           \colhead{$v$}&
           \colhead{Hour Angle} \\

           \colhead{($2,\!400,\!000$+)}&
           \colhead{Measured $V$}&
           \colhead{$\sigma_{V}$}&
           \colhead{Model $V$}&
           \colhead{$(O\!-\!C)_{V}$}&
           \colhead{(m)}&
           \colhead{(m)}&
           \colhead{(h)}}

\startdata

54237.763 & 0.864 & 0.086 & 0.783 & \phs0.081 & \phs202.4       & 250.7&  $-$2.24 \\
54237.774 & 0.909 & 0.107 & 0.775 & \phs0.134 & \phs196.7       & 258.2 &  $-$1.99 \\
54237.784 & 0.736 & 0.062 & 0.759 &  $-$0.022 & \phs190.3       & 265.2 &  $-$1.74 \\
54237.796 & 0.702 & 0.063 & 0.729 &  $-$0.027 & \phs182.4       & 272.6 &  $-$1.46 \\
54237.806 & 0.585 & 0.058 & 0.688 &  $-$0.103 & \phs174.6       & 278.9 &  $-$1.22 \\
54237.816 & 0.652 & 0.076 & 0.625 & \phs0.027 & \phs165.6       & 285.3 &  $-$0.97 \\
54237.833 & 0.468 & 0.053 & 0.474 &  $-$0.006 & \phs149.7       & 294.7 &  $-$0.56 \\
54237.932 & 0.833 & 0.049 & 0.833 & \phs0.001 & \phs\phn30.4    & 326.9 & \phs1.82 \\
54237.942 & 0.775 & 0.059 & 0.791 &  $-$0.017 & \phs\phn17.1    & 327.7 & \phs2.05 \\
54237.954 & 0.672 & 0.038 & 0.672 & \phs0.001 & \phs\phn\phn0.5 & 328.1 & \phs2.34 \\
54237.980 & 0.244 & 0.015 & 0.247 &  $-$0.004 & \phn$-$35.3     & 326.5 & \phs2.98 \\
54247.701 & 0.858 & 0.113 & 0.887 &  $-$0.029 & \phs159.9       & 214.9 &  $-$3.08 \\
54247.716 & 0.888 & 0.080 & 0.863 & \phs0.025 & \phs154.1       & 223.0 &  $-$2.73 \\
54247.729 & 0.824 & 0.083 & 0.785 & \phs0.040 & \phs147.5       & 230.2 &  $-$2.40 \\
54247.744 & 0.669 & 0.093 & 0.644 & \phs0.025 & \phs139.1       & 237.6 &  $-$2.05 \\
54247.761 & 0.435 & 0.058 & 0.430 & \phs0.005 & \phs128.1       & 245.6 &  $-$1.64 \\
54249.714 & 0.589 & 0.053 & 0.621 &  $-$0.032 & \phs152.1       & 225.3 &  $-$2.63 \\
54249.726 & 0.570 & 0.054 & 0.609 &  $-$0.039 & \phs146.6       & 231.1 &  $-$2.36 \\
54249.739 & 0.575 & 0.064 & 0.573 & \phs0.002 & \phs138.6       & 238.1 &  $-$2.03 \\
54249.751 & 0.594 & 0.063 & 0.524 & \phs0.070 & \phs131.3       & 243.5 &  $-$1.75 \\
54249.772 & 0.391 & 0.059 & 0.376 & \phs0.015 & \phs115.7       & 252.8 &  $-$1.24 \\
54310.716 & 0.616 & 0.062 & 0.526 & \phs0.090 & \phs\phn48.7    & 325.0 & \phs1.49 \\
54310.726 & 0.405 & 0.050 & 0.410 &  $-$0.005 & \phs\phn35.8    & 326.4 & \phs1.72 \\
54310.776 & 0.477 & 0.050 & 0.454 & \phs0.023 & \phn$-$31.5     & 326.8 & \phs2.91 \\
54310.786 & 0.558 & 0.054 & 0.619 &  $-$0.061 & \phn$-$45.5     & 325.4 & \phs3.16 \\
54310.797 & 0.870 & 0.100 & 0.745 & \phs0.125 & \phn$-$59.5     & 323.5 & \phs3.42 \\

\enddata

\end{deluxetable}

\clearpage

%%%%%%%%%%%%%%%%% TABLE: SB Orbital Elements %%%%%%%%%%%%%%%%%%%%

%\voffset30pt{}
\begin{deluxetable}{lccc}
\rotate 
\tabletypesize{\footnotesize}
\tablecaption{Spectroscopic Orbital Solutions for $\sigma^2$~CrB
\label{tab:SBele}}
\tablewidth{0pt}

\tablehead{%\vspace{-25pt} \\
           \colhead{Element}&
           \colhead{This Work}&
           \colhead{SR03 Velocities\tablenotemark{a}}&
           \colhead{SR03 Results}}

\startdata

\multicolumn{4}{l}{Orbital Elements:} \\
\phn\phn $P$ (days)                                 & 1.139791423 $\pm$ 0.000000080\tablenotemark{b}        & 1.139791423 $\pm$ 0.000000080\tablenotemark{b}        & 1.1397912 (adopted)  \\
\phn\phn $T$ (HJD-$2,\!400,\!000$)\tablenotemark{c} & $50,\!127.61845 \pm 0.00020$\tablenotemark{b}\phs\phs & $50,\!127.61845 \pm 0.00020$\tablenotemark{b}\phs\phs & $50,\!127.6248$\tablenotemark{d}\phn \\
\phn\phn $e$                                        & 0.0\tablenotemark{e}                                  & 0.0\tablenotemark{e}                                  & 0.0\tablenotemark{e}                 \\
\phn\phn $\omega$ (deg)                             & 0.0\tablenotemark{e}                                  & 0.0\tablenotemark{e}                                  & 0.0\tablenotemark{e}                 \\
\phn\phn $\gamma$ (\kms)                            & $-$13.03 $\pm$ 0.11\phn\phs                           & $-$12.58 $\pm$ 0.05\phn\phs                           & $-$12.3 $\pm$ 0.06\phs \\
\phn\phn $K_{\rm p}$ (\kms)                         & 61.25 $\pm$ 0.21\phn                                  & 61.31 $\pm$ 0.06\phn                                  & 61.34 $\pm$ 0.06\phn \\
\phn\phn $K_{\rm s}$ (\kms)                         & 63.89 $\pm$ 0.22\phn                                  & 62.90 $\pm$ 0.08\phn                                  & 62.91 $\pm$ 0.08\phn \\
\multicolumn{4}{l}{Derived quantities:} \\
\phn\phn $M_{\rm p}\sin^3 i$ (\msun)                & 0.11818 $\pm$ 0.00092                                 & 0.11461 $\pm$ 0.00032                                 & \phn\phn\phn0.1147   \\
\phn\phn $M_{\rm s}\sin^3 i$ (\msun)                & 0.11329 $\pm$ 0.00086                                 & 0.11170 $\pm$ 0.00027                                 & \phn\phn\phn0.1118   \\
\phn\phn $q \equiv M_{\rm s}/M_{\rm p}$             & 0.9586 $\pm$ 0.0047                                   & 0.9746 $\pm$ 0.0016                                   & 0.975 $\pm$ 0.002    \\
\phn\phn $a_{\rm p}\sin i$ (10$^6$ km)              & 0.9600 $\pm$ 0.0033                                   & 0.96085 $\pm$ 0.00097                                 & 0.96138 $\pm$ 0.00093 \\
\phn\phn $a_{\rm s}\sin i$ (10$^6$ km)              & 1.0014 $\pm$ 0.0035                                   & 0.98592 $\pm$ 0.00126                                 & 0.9861 $\pm$ 0.0012  \\
\phn\phn $a\sin i$ (\rsun)                          & 2.8181 $\pm$ 0.0068                                   & 2.7971 $\pm$ 0.0023                                   & 2.798 $\pm$ 0.002    \\
\multicolumn{4}{l}{Other quantities pertaining to the fit:} \\
\phn\phn $N_{\rm obs}$                              & 46                                                    & 217                                                   & 217                 \\
\phn\phn Time span (days)                           & 2610                                                  & 5.4                                                   & 5.4                 \\
\phn\phn $\sigma_{\rm p}$ (\kms)\tablenotemark{f}   & 1.04                                                  & \phn0.74                                              & \phn0.71            \\
\phn\phn $\sigma_{\rm s}$ (\kms)\tablenotemark{f}   & 1.10                                                  & \phn0.97                                              & \ldots                \\

\enddata

\tablenotetext{a}{Our orbital solution using SR03 velocities.}
\tablenotetext{b}{Determined using all published velocities (see \S\,\ref{sec:historical})}
\tablenotetext{c}{$T$ is the epoch of maximum primary velocity.}
\tablenotetext{d}{The value from SR03 has been shifted by an integer
number of cycles to the epoch derived in this work, for comparison
purposes.}  
\tablenotetext{e}{Circular orbit adopted.}
\tablenotetext{f}{RMS residual from the fit.}

\end{deluxetable}

\clearpage

%%%%%%%%%%%%%%%%% TABLE: Visual Orbit Elements %%%%%%%%%%%%%%%%%%%%

%\voffset30pt{}
\begin{deluxetable}{lc}
\tabletypesize{\footnotesize}
\tablecaption{Visual Orbit Solution for $\sigma^2$~CrB
\label{tab:VBele}}
\tablewidth{0pt}

\tablehead{%\vspace{-25pt} \\
           \colhead{Orbital Parameter\phn\phn\phn\phn\phn}&
           \colhead{Value}}

\startdata

\multicolumn{2}{l}{Adopted values:} \\
\phn\phn Period (days)                                         & 1.139791423 $\pm$ 0.000000080\tablenotemark{a} \\
\phn\phn $T_{\rm node}$ (HJD-$2,\!400,\!000$)\tablenotemark{b} & $50,\!127.04855 \pm 0.00020$\phs\phs\phs \\
\phn\phn $e$                                                   & 0.0\tablenotemark{c}             \\
\phn\phn $\omega$ (deg)                                        & 0.0\tablenotemark{c}             \\
\phn\phn $\theta_{\rm p}$ (mas)                                & 0.50 $\pm$ 0.05\tablenotemark{d} \\
\phn\phn $\theta_{\rm s}$ (mas)                                & 0.45 $\pm$ 0.05\tablenotemark{d} \\
\multicolumn{2}{l}{Visual orbit parameters:} \\
\phn\phn $\alpha$ (mas)                                        & 1.225 $\pm$ 0.013                \\
\phn\phn $i$ (deg)                                             & 28.08 $\pm$ 0.34\phn             \\
\phn\phn $\Omega$ (deg)                                        & 207.93 $\pm$ 0.67\tablenotemark{e}\phn  \\
\phn\phn $\Delta K'$                                           & 0.19 $\pm$ 0.19                  \\
\phn\phn Reduced $\chi^2$                                      & \phn0.61\tablenotemark{f}        \\
\enddata

\tablenotetext{a}{See \S\,\ref{sec:historical}.}
\tablenotetext{b}{This is the epoch of the ascending node, defined as
the epoch of maximum secondary velocity, and accordingly is one-half
period less than the value in Table~\ref{tab:SBele} (see
\S\,\ref{sec:visualorbit}).}
\tablenotetext{c}{Circular orbit adopted.}
\tablenotetext{d}{See \S\,\ref{sec:visualorbit}.}
\tablenotetext{e}{This value suffers from a 180\sdeg~ambiguity due to
the cosine term in Equation (\ref{eq:Vbin}).}
\tablenotetext{f}{The low $\chi^2$ indicates that our error estimates
for visibility are conservative.}
\end{deluxetable}

\clearpage

%%%%%%%%%%%%%%%%% TABLE: Physical Parameters %%%%%%%%%%%%%%%%%%%%

%\voffset30pt{}
\begin{deluxetable}{lcccccc}
\rotate 
\tabletypesize{\footnotesize}
\tablecaption{Physical parameters for $\sigma^2$~CrB
\label{tab:PhyPar}}
\tablewidth{0pt}

\tablehead{%\vspace{-25pt} \\
           \colhead{}&
           \multicolumn{2}{c}{This Work}&
           \multicolumn{2}{c}{SR03 Spectroscopy\tablenotemark{a}}&
           \multicolumn{2}{c}{SR03 Results} \\

           \colhead{Physical Parameter}&
           \colhead{Primary}&
           \colhead{Secondary}&
           \colhead{Primary}&
           \colhead{Secondary}&
           \colhead{Primary}&
           \colhead{Secondary}}

\startdata

$a$ (\rsun)        & \multicolumn{2}{c}{5.99 $\pm$ 0.07}     & \multicolumn{2}{c}{5.99 $\pm$ 0.07}           & \multicolumn{2}{c}{\ldots} \\
Mass (\msun)       & 1.137 $\pm$ 0.037  & 1.090 $\pm$ 0.036  & 1.128 $\pm$ 0.037     & 1.099 $\pm$ 0.036     & 1.108 $\pm$ 0.004\tablenotemark{b} & 1.080 $\pm$ 0.004\tablenotemark{b} \\
Radius (\rsun)     & 1.244 $\pm$ 0.050  & 1.244 $\pm$ 0.050  & 1.244 $\pm$ 0.050     & 1.244 $\pm$ 0.050     & 1.14 $\pm$ 0.04                    & 1.14 $\pm$ 0.04  \\
$T_{\rm eff}$ (K)  & 6050 $\pm$ 150\phn & 5870 $\pm$ 150\phn & 6000 $\pm$ 50\phn\phn & 5900 $\pm$ 50\phn\phn & 6000 $\pm$ 50\phn\phn              & 5900 $\pm$ 50\phn\phn \\
$M_{\rm V}$ (mag)  & 4.35 $\pm$ 0.02    & 4.74 $\pm$ 0.02    & 4.45 $\pm$ 0.02       & 4.61 $\pm$ 0.02       & 4.61 $\pm$ 0.07                    & 4.76 $\pm$ 0.07 \\
$M_{\rm K}$ (mag)  & 2.93 $\pm$ 0.09    & 3.12 $\pm$ 0.11    & \ldots                & \dots                 & \ldots                             & \ldots \\

\enddata

\tablenotetext{a}{These parameters use the SR03 spectroscopic results
such as flux ratio, rotational velocities, and radial velocities, but
use the \citet{Les1999} parallax, Tycho-2 magnitudes, and our
visual orbit.}
\tablenotetext{b}{As noted in \S\,\ref{sec:introduction}, these
uncertainties are unrealistically small.}

\end{deluxetable}

\clearpage

%%%%%%%%%%%%%%%%% TABLE: Wide Visual Orbit (AB) %%%%%%%%%%%%%%%%%%%%

%\voffset30pt{}
\begin{deluxetable}{lc}
\tabletypesize{\footnotesize}
\tablecaption{Visual Orbit Solution for $\sigma^1\!-\!\sigma^2$~CrB 
\label{tab:WideVis}}
\tablewidth{0pt}

\tablehead{%\vspace{-25pt} \\
           \colhead{Orbital Parameter}&
           \colhead{Value}}

\startdata

$P$ (years)        & 726           $\pm$ 62\phn     \\
$T_{\rm 0}$ (BY)   & $1,\!825.2 \pm 1.5$\phn\phn\phn    \\
$e$                & 0.72          $\pm$ 0.01       \\
$\omega$ (deg)     & 237.3         $\pm$ 6.8\phn\phn \\
$\alpha$ (arcsec)  & 5.26          $\pm$ 0.35       \\
$i$ (deg)          & 32.3          $\pm$ 4.1\phn    \\
$\Omega$ (deg)     & 28.0          $\pm$ 0.5\phn    \\

\enddata

\end{deluxetable}

\clearpage

%%%%%%%%%%%%%%%%%%%%%%%%%%%%%%%%%%% FIGURES %%%%%%%%%%%%%%%%%%%%%%%%%%%%%%%%%%

\clearpage

\clearpage

\clearpage

\clearpage

\clearpage

\clearpage

\clearpage

\clearpage

\clearpage

\clearpage

\clearpage

\clearpage

\clearpage

\clearpage

%%%%%%%%%%%%%%%%%%%%%%%%%%%%%%%%%%%%%%%%%%%%%%%%%%%%%%%%%%%%%%%%%%%%%%%%%%%%%%
%%%%%%%%%%%%%%%%%%%%%%%%%%%%%%%%% THE END %%%%%%%%%%%%%%%%%%%%%%%%%%%%%%%%%%%%
%%%%%%%%%%%%%%%%%%%%%%%%%%%%%%%%%%%%%%%%%%%%%%%%%%%%%%%%%%%%%%%%%%%%%%%%%%%%%%
\end{document}